\documentclass[]{article}
\usepackage{amssymb}

\newcommand\mathC{{\mkern1mu\raise2.2pt\hbox{
$\scriptscriptstyle|$} {\mkern-7mu\rm C}}}
\newcommand{\mathR}{{\rm I\! R}}% The real number system
\renewcommand\[{[\,}            %% Redefinition!
\renewcommand\]{\,]}            %% Redefinition!

\renewcommand\mathR{{\rm I\! R}}

\newcommand\unit{{\rm 1\kern-3.2pt I}}

\newcommand\g{\gamma}

\newcommand\z{\zeta}
\renewcommand\l{\lambda}        %% Redefinition!

\newcommand{\R}{{\rm I\! R}}       % The real number system

\begin{document}
\title{Mini-superspace models in Histories Theory}
\author{Charis Anastopoulos\footnote{anastop@physics.upatras.gr}\\
{\small Department of Physics, University of Patras, 26500 Patras,
Greece}
\\
 and \\ Ntina Savvidou \footnote{ntina@imperial.ac.uk} \\
{\small Department of Physics, University of Patras, 26500 Patras,
Greece} \\ {\small and} \\
{\small  Theoretical Physics Group, Imperial College, SW7 2BZ,
London, UK}}

\maketitle

\renewcommand {\thesection}{\arabic{section}}
 \renewcommand {\theequation}{\thesection. \arabic{equation}}
\let \ssection = \section
\renewcommand{\section}{\setcounter{equation}{0} \ssection}

\begin{abstract}
We study the Robertson-Walker minisuperspace model in histories
theory, motivated by the  results that emerged from the histories
approach to general relativity. We examine, in particular, the
issue of time-reparameterisation in such systems. The model is
quantised using an adaptation of reduced-state-space quantisation.
We finally discuss the classical limit, the implementation of
initial cosmological conditions and estimation of probabilities in
the histories context.
\end{abstract}

\section{Introduction}

The Robertson-Walker model provides the standard paradigm for
modern cosmology since it incorporates the symmetries of spatial
homogeneity and isotropy in the solution of Einstein equations. In
cosmology the R-W model appears as a classical dynamical system,
that it is described by the scale factor and by spatially averaged
matter field variables.

\bigskip

In spite of the success of the R-W model at the classical level,
there has been a strong effort to study the quantisation of this,
as well as other cosmological models, motivated by their role in
the search for a quantum gravity theory.

Such models allow us to probe the early---Planck length---stages
of the Universe. In spite of the absence of a full theory of
quantum gravity they allow a testing ground for hypotheses about
the initial condition of the Universe, the emergence of classical
spacetime and the plausibility of the special initial conditions
for inflation.

 'Minisuperspace' models are very simple systems, as they have been
divested of much of the symmetry of general relativity, still they
provide non-trivial examples in which quantum gravity programmes
may apply their ideas and techniques. They are particularly
relevant for the discussion of major conceptual issues of quantum
gravity such as the problem of time, the construction of
reparameterisation-invariant physical observables and the physical
interpretation of the Hamiltonian constraint.

\bigskip

In this work we study the quantisation of minisuperspace models
within the context of the Histories Projection Operator formalism,
in light of the significant results emerged from the histories
approach to general relativity\cite{Sav03a, Sav03b}.

The (HPO) histories formalism is a continuation of the consistent
histories theory developed by Griffiths, Omn\'es, Gell-Mann and
Hartle---for reviews see \cite{Gri03, Omn94, Omn99, GeHa90,
Har93a}. The fundamental object in this formalism is the {\em
history}, which is defined as a time-ordered sequence of
propositions about properties of a physical system. When a certain
`decoherence condition' is satisfied by a set of histories then
the elements of this set {\em can\/} be given probabilities. The
probability information of the theory is encoded in the
decoherence functional, a complex function of pairs of histories.

The importance of the histories scheme for quantum gravity is that
histories are objects that are intrinsically temporal, namely they
incorporate in their definition the notion of causal ordering.
This is very desirable for a theory of quantising gravity, where
the loss of time at the space of true degrees of freedom---the
problem of time \cite{I92, Kuc91}---is considered as one of the
main conceptual problems. Indeed, the development of the histories
formalism in the Histories Projection Operator scheme \cite{I94,
IL94, IL95, ILSS98, Sav99, Sav01b}, has led to a quantisation
algorithm for parameterised systems \cite{SavAn00} in which the
histories causal ordering is preserved on the space of true
degrees of freedom.

Of great importance in the development of the histories scheme is
the construction of a classical  histories theory \cite{Sav99}.
The basic mathematical entity is the space of differentiable paths
taking their value in the space $\Gamma$ of classical states. The
key idea in this new approach to classical histories is contained
in the symplectic structure on this space of temporal paths.
Analogous to the quantum case, there are generators for two types
of time transformation: one associated with classical temporal
logic, and one with classical dynamics. One significant feature is
that the paths corresponding to solutions of the classical
equations of motion are determined by the requirement that they
remain invariant {\em under the symplectic transformations
generated by the action\/}\cite{Sav99}.

The strongest motivation for the study of minisuperspace models
within the histories theory comes from its application at the
level of general relativity from where two important results
arise. First, the spacetime diffeomorphism group coexists with the
`group' generated by the history version of the canonical
constraints \cite{Sav01, Sav03a}. The constraints, however, depend
on a choice of foliation for the 3+1 splitting of spacetime. This
leads to the important question whether physical results depend
upon this choice. The solution of the constraints determines a
reduced phase space for histories, which has an explicit
dependence on the foliation. It turns out that {\em space-time
diffeomorphisms intertwine between different such
reductions\/}\cite{Sav03b}. This is a completely novel result,
which has been made possible only by the incorporation of general
relativity into the histories formalism.

The results above suggest that a scheme for the quantisation of
gravity based on histories has two highly desirable
features\cite{Sav03b}.

First, the Lorentzian metric can be directly quantised; this
contrasts with conventional canonical quantum gravity where only a
spatial three-metric is quantised.

Second, the history scheme incorporates general covariance via a
manifest representation of the space-time diffeomorphism group.

\bigskip

Even though  minisuperspace models lack the most fundamental
feature of general relativity---general covariance---, they still
provide an important testing ground of ideas and techniques for a
quantum gravity programme. The construction of a Hilbert space of
histories, the implementation of the constraints  and the
identification of gauge invariant physical observables are highly
non-trivial procedures even in the simplest of minisuperspace
models--for other studies of minisuperspace models in the
framework of consistent histories see \cite{Hal} . As such they
sharpen our understanding about the issues involved and allow us
to distinguish between the deep conceptual problems and the ones
that are of a purely technical nature.

The plan of the paper is as follows. In the next section we
provide a concise description of the classical histories theory,
emphasising the insights obtained by the incorporation of general
relativity into the histories programme.

In section 3, we study the classical histories theory for an R-W
cosmological model with a scalar field. We identify the histories
reduced state space for various values of the relevant parameters.
We place strong emphasis in the notion of time-reparameterisations
for this system, as it is the only remnant of the spacetime
diffeomorphism invariance in the minisuperspace model. We note
that in some cases the histories reduced state space has the
structure of an orbifold---it is not a globally smooth manifold.

In section 4, we describe the construction of the history Hilbert
space and the decoherence functional for cosmological models.
Starting from previous results we discuss the issue of finding a
representation of the theory when the Hamiltonian does not possess
a vacuum state. Such is the case of minisuperspace models for
which the Hamiltonian constraint does not have a continuous
spectrum around zero. We analyse the difference in physical
predictions entailed by the choice of different representations.

An important feature of the construction of the histories Hilbert
space is that it strongly relies on the identification of a set of
canonical coherent states. For this reason we analyse the
implementation of the constraints at the coherent states level.

In section 5, after a brief discussion of Dirac quantisation in
the histories setting \cite{SavAn00}, we suggest that the most
adequate method of implementing the constraints at the quantum
level is a version of reduced state space quantisation. We analyse
this procedure in detail and we construct the canonical coherent
states for the reduced system.

In section 6 we explicitly construct the full history Hilbert
space, and identify the physical history Hilbert space. We then
write the decoherence functional for the true degrees of freedom.
We expand on the issue of the initial condition and the emergence
of classical spacetime. We examine how the 'tunnelling' initial
condition \cite{Vil84} may be implemented in the histories scheme,
providing an estimation of the tunnelling probability. Finally we
discuss the extensions of the formalism involved in a rigorous
treatment of the non-boundary proposal \cite{HaHa83} in the
histories framework.

\section{Background}

\subsection{Classical Histories}

In the consistent histories formalism,  a history $ \alpha = (
\hat{\alpha}_{t_{1}},\hat{\alpha}_{t_{2}},\ldots,
\hat{\alpha}_{t_{n}})$ is defined to be a collection of projection
operators $\hat\alpha_{t_i}$, $i=1,2,\ldots,n$, each of which
represents a property of the system at the single time $t_i$.
Therefore, the emphasis is placed on histories, rather than
properties at a single time, which in turn gives rise to the
possibility of generalized histories with novel concepts of time.

The HPO approach, places particular emphasis on temporal logic.
This is achieved by representing the history $\alpha$ as the
operator $\hat\alpha :=\hat{\alpha}_{t_{1}}\otimes\hat{\alpha}
_{t_{2}}\otimes\cdots \otimes \hat{\alpha}_{t_{n}}$ which is a
genuine {\em projection\/} operator on the tensor product $
\otimes_{i=1}^n {\mathcal{H}}_{t_i} $ of copies of the standard
Hilbert space $\mathcal{H}$.

 The space of classical histories $
\Pi = \{ \gamma \mid \gamma : \mathR \rightarrow \Gamma \}$ is the
set of all smooth paths on the classical state space $\Gamma$. It
can be equipped with a natural symplectic structure, which gives
rise to the Poisson bracket. For the simple case of a particle at
a line
\begin{eqnarray}
    \{x_t\, , x_{t^{\prime}} \}_{\Pi} &=& 0\\
    \{p_t\, , p_{t^{\prime}} \}_{\Pi} &=& 0\\
  \{x_t\, , p_{t^{\prime}} \}_{\Pi} &=& \delta (t-t^{\prime})
  \end{eqnarray}
where
\begin{eqnarray}
x_t : \Pi & \rightarrow & \mathR  \\
\gamma & \mapsto & x_t (\gamma) := x( \gamma (t))
\end{eqnarray}
and similarly for $ p_t $.

One may define the Liouville function
 \begin{equation}
  V(\gamma) := \int_{-\infty}^\infty \! dt\, p_t\, \dot{x}_t ,
 \end{equation}
which generates symplectic transformations of the form $(x_t,p_t)
\rightarrow (x_{t+s}, p_{t+s})$; also the Hamiltonian ({\em i.e.},
time-averaged energy) function $H$
\begin{equation}
  H ( \gamma ):= \int_{-\infty}^{\infty}\!dt \,h_t(x_t,p_t)
\end{equation}
where $h_t$ is the Hamiltonian that is associated with the copy
${\Gamma}_t$ of the normal classical state space with the same
time label $t$.

The history equivalent of the classical equations of motion is
given by the following condition\cite{Sav99} that holds for
\textit{all} functions $F$ on $\Pi$ when $\gamma_{cl}$ is a
classical solution:
\begin{equation}
\{F , S\}_{\Pi}\, (\gamma_{cl}) = 0, \label{eqnsofmotn}
\end{equation}
where
\begin{equation}
S( \gamma ) := \int_{-\infty}^{\infty}\!dt \, (p_t\dot{x_t} -
H_t(x_t,p_t))=\, V( \gamma )- H( \gamma )
\end{equation}
is the classical analogue of the action operator \cite{Sav99}.

\paragraph*{Classical parameterised systems.}
Parameterised systems have a vanishing Hamiltonian $H = h(x , p)
$, when the constraints are imposed. Classically, two points of
the constraint surface $C$ correspond to the same physical state
if they are related by a transformation generated by the
constraint. The true degrees of freedom correspond to equivalence
classes of such points and are represented by points of the
reduced state space ${\Gamma}_{red}$.

In the histories approach to parameterised systems, the history
constraint surface $C_h$ is defined as the set of all smooth paths
from the real line to the constraint surface $C$. The history
Hamiltonian constraint is defined by $H_{\kappa} = \int \! dt\,
\kappa (t) h_t$, where $h_t := h(x_t, p_t)$ is first-class
constraint. For all values of the smearing function $\kappa (t)$,
the history Hamiltonian constraint $H_{\kappa}$ generates
canonical transformations on the history constraint surface $C_h$.
The history reduced state space $\Pi_{red}$ is then defined as the
set of all smooth paths on the canonical reduced state space
$\Gamma_{red}$, and it is identical to the space of orbits of
$H_{\kappa}$ on $C_h$.

In order for a function on the full state space, $ \Pi $, to be a
physical observable ({\em i.e.,\/} to be projectable into a
function on $ {\Pi}_{red}$), it is necessary and sufficient that
it commutes with the constraints on the constraint surface.

Contrary to the canonical treatments of parameterised systems, the
classical equations of motion are explicitly realised on the
reduced state space $\Pi_{red}$. They are given by
\begin{equation}
\{ \tilde S , F \}\,(\gamma_{cl}) = \{ \tilde{V} , F
\}\,(\gamma_{cl}) = 0
\end{equation}
where $\tilde{S}$ and $\tilde{V}$ are respectively the action and
Liouville functions projected on $\Pi_{red}$. Both $\tilde{S}$ and
$\tilde{V}$ commute weakly with the Hamiltonian constraint
\cite{SavAn00}.

\subsection{Histories general relativity}
We will next describe the incorporation of classical general
relativity into the histories framework \cite{Sav03a, Sav03b}.

We consider a four-dimensional manifold $M$, which has the
topology $\R \times \Sigma$. The history space is defined as
$\Pi^{cov} = T^{*}{\rm LRiem}(M)$, where ${\rm LRiem}(M)$ is the
space of all Lorentzian four-metrics ${g_{\mu\nu}}$, and
$T^{*}{\rm LRiem}(M)$ is its cotangent bundle. Hence, the history
space $\Pi^{cov}$ for general relativity is the space of all
histories $(g_{\mu\nu}, \pi^{\mu\nu})$.

The history space $\Pi^{cov}$ is equipped with the symplectic form
\begin{equation}
 \Omega = \int \!d^4X \, \delta \pi^{\mu\nu}(X) \wedge \delta
 g_{\mu\nu}(X) \,, \label{omega}
\end{equation}
where $X$ is a point in the spacetime $M$, and where
$g_{\mu\nu}(X)$ is a four-metric that belongs to the space of
Lorentzian metrics $L {\rm Riem}(M)$, and $\pi^{\mu\nu}(X)$ is the
conjugate variable.

The symplectic structure Eq.\ (\ref{omega}) generates the
following covariant Poisson brackets algebra, on the history space
$\Pi^{cov}$
\begin{eqnarray}
\{g_{\mu\nu}(X)\,, \,g_{\alpha\beta}(X^{\prime})\}&=& 0
\label{covgg}\\
\{\pi^{\mu\nu}(X)\,, \,\pi^{\alpha\beta}(X^{\prime})\} &=& 0
\label{covpipi} \\
\{g_{\mu\nu}(X)\,, \,\pi^{\alpha\beta}(X^{\prime})\} &=&
\delta_{(\mu\nu )}^{\alpha\beta} \,\delta^4 (X, X^{\prime}) ,
\label{covgpi}
\end{eqnarray}
 where  ${{\delta}_{(\mu\nu)}}^{\alpha\beta} :=
\frac{1}{2}({\delta_\mu}^\alpha {\delta_\nu}^\beta +
{\delta_\mu}^\beta {\delta_\nu}^\alpha )$.

We  define the Liouville function $V_W$ associated with any vector
field $W$ on $M$ as
\begin{equation}
V_W:=\int \!d^4X \,\pi^{\mu\nu}(X)\,{\cal L}_W g_{\mu\nu}(X)
\label{Vw}
\end{equation}
where ${\cal L}_W$ denotes the Lie derivative with respect to $W$.

These  functions $V_W$, defined for any vector field $W$ as in
Eq.\ (\ref{Vw}), satisfy the Lie algebra of the {\em spacetime
diffeomorphisms group\/} ${\rm Diff}(M)$:
\begin{equation}
\{\, V_{W_1}\,, V_{W_2}\,\} = V_{ [ W_1 , W_2 ]},
\end{equation}
where $[ W_1 , W_2 ]$ is the Lie bracket between vector fields
$W_1$ and $W_2$ on the manifold $M$.

We  next introduce a $3+1$ foliation of the spacetime $M$, which
is spacelike with respect to a Lorentzian metric $g$, in order to
construct a $3 + 1$ description of the theory. However, a key
feature of the present construction is that this foliation is
required to be {\em four-metric dependent\/} in order to address
the key issue of requiring all the different choices of foliation
to be spacelike.

For each $g \in LRiem(M)$ we choose a spacelike foliation ${\cal
E}[g]$. For a given Lorentzian metric $g$, we use the foliation
${\cal{E}}[g]$ to split $g$ with respect to the Riemannian
three-metric $h_{ij}$, the lapse function $N$ and the shift vector
$N^i$ as
\begin{eqnarray}
 h_{ij}(t,\underline{x};g] &:=&
 {\cal{E}}^{\mu}_{,i}(t,\underline{x};g]\,
 {\cal{E}}^{\nu}_{,j}(t,\underline{x};g]\,
 g_{\mu\nu}({\cal{E}}(t,\underline{x};g]) \label{Pullbackh[g]} \\
 N_{i}(t,\underline{x};g] &:=&
 {\cal{E}}^{\mu}_{,i}(t,\underline{x};g]\,
 {\dot{\cal{E}}}^{\nu}(t,\underline{x};g]\, g_{\mu\nu}({\cal{E}}
 (t,\underline{x};g])\label{shift[g]} \\
 -N^{2}(t,\underline{x};g] &:=&
 {\dot{\cal{E}}}^{\mu}(t,\underline{x};g]\,
 {\dot{\cal{E}}}^{\nu}(t,\underline{x};g]\,
 g_{\mu\nu}({\cal{E}}(t,\underline{x};g]) - N_{i} N^{i} (t,
 \underline{x})
 \label{lapse[g]}
\end{eqnarray}

Detailed calculations show that the symplectic form $\Omega$ can
be written in the equivalent canonical form, with respect to a
chosen foliation functional $\cal E$
\begin{eqnarray}
 \Omega &=& \int  \int\!d^3x\,dt (\delta \pi^{ij}\wedge\delta h_{ij} +
 \delta p \wedge\delta N
 + \delta p_{i}\wedge \delta N^{i}), \label{omegacan}
\end{eqnarray}
where the momenta have been suitably defined in terms of
$\pi^{\mu\nu}$ and the foliation-functional.

One then define the history version of the canonical constraints
which satisfy a histories version of the Dirac algebra
\cite{Sav01, Sav03a}.

\bigskip

To summarise, the relation between the spacetime diffeomorphism
algebra, and the Dirac constraint algebra has long been an
important matter for discussion in quantum gravity. Therefore it
is of great significance that, in this new construction, the two
algebras appear together in an explicit way: the classical theory
contains realisations of both the space-time diffeomorphism group
{\em and\/} the Dirac algebra.

Furthermore the canonical constraints can be straightforwardly
written using the paths of canonical metric and momenta. This is a
very significant result: it implies that there is a central role
for {\em spacetime\/} concepts, as opposed to the domination by
spatial ideas in the normal canonical approaches to quantum
gravity.

The constraints, however, depend on the foliation functional. This
leads to the important question whether physical results depend
upon this choice. The solution of the constraints determines a
reduced phase space for histories, which has an explicit
dependence on the foliation. It is proved the critical result that
{\em space-time diffeomorphisms intertwine between different such
reductions\/}. This is a completely novel result, which has been
made possible only by the incorporation of general relativity into
the histories formalism.

In particular. the spacetime diffeomorphisms generated by $V_W$
commute with the constraints if the foliation functional satisfies
the equivariance condition: For a function ${\cal E}:{\rm
LRiem}(M)\rightarrow {\rm Fol}(M)$, (where ${\rm LRiem}(M)$ is the
space of Lorentzian metrics on $M$, and ${\rm Fol}(M)$ is the
space of foliations of $M$) we say that ${\cal E}$ is an
`equivariant foliation' if
\begin{equation}
{\cal E}(f^*g)=f^{-1}\circ {\cal E}(g),
\label{Def:EquivariantFoliation2}
\end{equation}
for all Lorentzian metrics $g$ and spacetime diffeomorphisms
$f\in{\rm Diff}(M)$.

\section{Classical histories minisuperspace model}

\subsection{The Robertson-Walker model}
We study the reduction of the full general relativity histories
theory to a homogeneous and isotropic cosmological model with a
scalar field.

Our starting point is a configuration space containing the
Lorentzian metric $g_{\mu \nu}(X)$ and the scalar field $\Phi(X)$.
The history state space $\Pi_{cov}$ is the cotangent bundle over
the configuration space $\Pi_{RW} = T^*[LRiem(M) \times
C^{\infty}(M)]$. It is equipped with the symplectic form, in full
analogy with Eq.\ (\ref{omega}) of the symplectic form of general
relativity,
\begin{eqnarray}
\Omega = \int \!d^4X \, [\delta \pi^{\mu\nu}(X) \wedge \delta
 g_{\mu\nu}(X) + \delta \pi(X)\wedge \delta \Phi(X).
 \label{omega2}
\end{eqnarray}

Next we restrict the configuration space to a submanifold
consisting of metrics and scalar fields  of the form
\begin{eqnarray}
ds^2 &=& - A^2(X^0) (dX\!^0)^2 + e^{2a(X^0)} d{\Omega_3}^2 \\
\Phi(X) &=& \Phi(X^0),
\end{eqnarray}
where $d {\Omega_3}^2 = h^{\kappa}_{ij}\,\,dX^i dX\!^j$ and
$h^{\kappa}_{ij}$ is the homogeneous metric of constant curvature
equal to $\kappa \in \{-1,0, 1\}$ on the surfaces $X^0 =
constant$. The variable $A^2$ is the $00$ component of the metric
and $e^a$ is the scale factor of the R-W universe.

Substituting into Eq.\ (\ref{omega2}) we obtain the reduced
symplectic form
\begin{eqnarray}
\Omega = \int\! dX\!^0 [\delta \pi_A(X\!^0)\!\wedge\! \delta
A(X\!^0) + \delta \pi_{\alpha}(X\!^0) \!\wedge\! \delta a(X\!^0) +
\delta \pi(X\!^0) \!\wedge\! \delta \Phi(X\!^0)],
\label{omegaminicov}
\end{eqnarray}
where
\begin{eqnarray}
\pi_a(X^0) :&=& 2 \int d^3X \pi^{ij}(X) h_{ij}^{\kappa}(X^i)
e^{2a(X^0)} \\
\pi_A(X^0)) :& = & - 2 \int d^3X \pi^{00}(X)  A(X^0) \\
\pi_{\phi}(X^0) :&=& \int d^3 \pi_{\phi}(X).
\end{eqnarray}

The history state space $\Pi_{RW}$ for the R-W model is the
cotangent bundle of the space of paths in the variables $A, a,
\phi$, namely maps from $\R$ to $\R^3$, such  $X^0 \rightarrow
[A(X^0), a(X^0), \phi(X^0)]$. The symplectic form Eq.\
(\ref{omegaminicov}) is non-degenerate on the state space
$\Pi_{RW}$.

The spacetime diffeomorphisms are also projected on to the history
space of the reduced model albeit they reduce to diffeomorphisms
of the real line $\R$. They are generated by functions of the form
\begin{eqnarray}
V_W = \int dX^0 [ \pi_a {\cal L}_Wa + \pi_A {\cal L}_W A +
\pi_{\phi} {\cal L}_W \phi] \label{minidiff},
\end{eqnarray}
for any vector field $W = W(X^0) \frac{\partial}{\partial X^0}$.

The symplectic form Eq.\ (\ref{omegaminicov}) is written in a
covariant or spacetime way. In order to write its canonical
expression one needs to introduce a spacetime foliation.

In the case of histories general relativity, the introduction of
metric-dependent foliations guaranteed the spacetime character of
the canonical description, namely that the pull-back of the
Lorentzian metric on the foliation three-surfaces is always a
Riemannian metric\cite{Sav03a, Sav03b}. In the present context,
however, the restriction into a homogeneous and isotropic
spacetime has already taken care of this issue. While it is
possible to introduce a metric-dependent foliation however there
is no overriding physical reason demanding its introduction.

In the present model the definition of a foliation is equivalent
to a choice of a time parameter $t$. In other words we introduce
the function ${\cal E}: \R \rightarrow \R$, such that $t
\rightarrow X^0 = {\cal E}(t)$. ${\cal E}$ effects
time-reparameterisations and for this reason it is a strictly
increasing bijective function\footnote{If we were to choose to a
metric- dependent foliation ${\cal E}$, it should also be a
functional of the components of the metric and it would read
${\cal E}(t,a(\cdot),A(\cdot)]$.}.

We may then define the canonical configuration variables
\begin{eqnarray}
\l_t &:=& a(\,{\cal E}(t,a(\cdot),A(\cdot)]\,) \\
N_t &:=& A(\,{\cal E}(t,a(\cdot),A(\cdot)]\,) \\
\phi_t &:=& \Phi(\,{\cal E}(t,a(\cdot),A(\cdot)]\,)
\end{eqnarray}

The symplectic form Eq.\ (\ref{omegaminicov}) then becomes
\begin{eqnarray}
\Omega = \int dt [\delta \pi_{\l t} \wedge \delta \l_t + \delta
\pi_{N t} \wedge \delta N_t + \delta \pi_{\phi t} \wedge \delta
\phi_t],  \label{omegamincan}
\end{eqnarray}
where we have introduced the variables
\begin{eqnarray}
\pi_{\l t} &:=& \dot{{\cal E}}(t)\int dt' G_{\l}(t,t')
\pi_a({\cal E}(t')) \\
\pi_{N t} &:=& \dot{{\cal E}}(t)\int dt' G_{N}(t,t')
\pi_N({\cal E}(t')) \\
\pi_{\phi t} &:=& \dot{{\cal E}}(t) \int dt' G_{\phi}(t,t')
\pi_{\Phi}({\cal E}(t')),
\end{eqnarray}
and where $G_{\l}(t,t'), G_{N}(t,t'), G_{\phi}(t,t')$ are complex
kernels that depend on the functional derivatives $\frac{\delta
\cal{E}}{\delta a(t)}$ and $\frac{\delta \cal{E}}{\delta N(t)}$
\cite{Sav03a, Sav03b}. In the case of non-metric-dependent
foliations they reduce to delta functions. Note that the momentum
variables are scalar densities of weight 1, with respect to time.

We should remark here that Eq.\ (\ref{omegamincan}) can be
alternatively obtained from the canonical histories symplectic
form, by substituting a reduced expression for the metric, the
lapse function and the shift vector. However we follow this
approach in order to demonstrate the existence of a representation
of the diffeomorphisms on the real line---see section 3.3.

\subsection{Constraints}
Next we follow the standard canonical analysis of minisuperspace
models, hence we identify the constraints
\begin{eqnarray}
C&=& \frac{1}{2}N \left(\frac{1}{\sqrt{h}}(-\pi_{\l}^2 +
\pi_{\phi}^2) + \sqrt{h}
(V(\phi) - \kappa e^{\l}) \right) = 0\\
\pi_{N} &=& 0,
\end{eqnarray}
that they form a first-class system. Here $V(\phi)$ is a potential
for the scalar field $\phi$ and $h = e^{6 \l}$ is the determinant
of the three-metric. It is important to remark that $\sqrt{h}$ is
a scalar density of weight one with respect to {\em time}, since
$\sqrt{-g} = N \sqrt{h}$ and $\sqrt{-g}$ is a spacetime density of
weight one and $N$ is a scalar function\cite{Sav03a}. Even though
$\sqrt{h}$ is numerically equal to $e^{3 \l}$ it transforms
differently under diffeomorphisms of the real line and for this
reason we make this distinction in the expression for the
constraint $h$\footnote{This point is important because the
histories constraints involve an integration over time and one has
to keep an account of the densities for the definition to be
independent of the choice of the $t$ coordinate. This is not a
feature of the histories formalism alone; it also appears in the
construction of the phase space canonical action, which is an
object that appears also in the standard formalism.}.

The histories version of the canonical constraints are
\begin{eqnarray}
\hspace*{-0.6cm} C(\xi)\!\!\! &=& \!\!\frac{1}{2}\int\! dt \xi(t)
N \left(\frac{1}{\sqrt{h_t}}(-\pi_{\l t}^2 + \pi_{\phi t}^2) +
\sqrt{h_t} (V(\phi_t) - \kappa e^{\l_t}
\right) = 0 \\
\hspace*{-0.6cm} \pi_{u}\!\!\!&=&\!\! \int\! dt u(t) \pi_{N t}\,\,
= 0,
\end{eqnarray}
for all scalar smearing functions $\xi(t)$ and $u(t)$.

An alternative characterisation of the constraints involves the
introduction of an arbitrary scalar density $w(t)$, of weight 1.
We redefine the momenta
\begin{eqnarray}
p_{\l t} &=& \pi_{\l t}/w(t), \\
p_{N t} &=& \pi_{N t} / w(t), \\
p_{\phi t} &=& \pi_{\phi t}/w(t),
\end{eqnarray}
and we rescale the metric determinant $\sqrt{h'} := \sqrt{h}
/w(t)$, so that $p_{\l t}, p_{N t}, p_{\phi t}, $ and
$\sqrt{h'_t}$ are all scalar functions with respect to time. The
Hamiltonian constraint is then written
\begin{eqnarray}
C_{\xi} =\frac{1}{2}\int dt \xi(t) w(t) e^{-3\lambda_t} [ -p_{\l
t}^2 + p_{\phi t}^2 + V(\phi_t) e^{6 \l_t}  - \kappa e^{4 \l_t}].
\end{eqnarray}
The price to pay then is that the symplectic form, (hence the
Poisson brackets), involves explicitly the weight function $w(t)$
\begin{eqnarray}
\Omega = \int dt w(t) [\delta p_{\l t} \wedge \delta \l_t + \delta
p_{N t} \wedge \delta N_t + \delta p_{\phi t} \wedge \delta
\phi_t].
\end{eqnarray}

The state space action functional for the model is written
\begin{eqnarray}
S = V -H_N - \pi_N(u),
\end{eqnarray}
in terms of the Liouville function $V$, that generates time
translations
\begin{eqnarray}
V = \int dt (\pi_{\l t} \dot{\l}_{t} + \pi_{Nt} \dot{N}_t +
\pi_{\phi t} \dot{\phi}_t).
\end{eqnarray}

The next step involves the implementation of the constraints and
the construction of the reduced histories state space. The primary
constraint $\pi_{N t} = 0 $ is trivially implemented. To implement
the Hamiltonian constraint we first need to solve the constraint
$H_t = 0$.

To simplify our analysis we shall assume that the scalar field
potential $V(\phi)$ is a constant and it is equal to the
cosmological constant $\Lambda > 0$. The constraint equation then
reads $ -p_{\l t}^2 + p_{\phi t}^2 + U(\l_t0 = 0$, where $U(\l_t)
= e^{6 \l_t} - \kappa e^{4 \l_t}$ plays the role of a 'potential'
for the variable $\l$. There exist then two physically distinct
cases.

\subsubsection{ $\Lambda > 0, \kappa \in \{0,-1\}$}
 In
this case $U(\l_t) > 0$ for all values of $\l_t$ and we may solve
the constraint with respect to $\pi_{\l t}$
\begin{eqnarray}
\pi_{\l t} = \pm \sqrt{\pi_{\phi t}^2 + U(\l_t)}. \label{solve}
\end{eqnarray}

The constraint surface consists of two disconnected components,
one for each sign of $\pi_{\l}$ in the right-hand-side of equation
Eq.\ (\ref{solve}). The coordinate expression for the symplectic
form $\Omega_C$, restricted on the history constraint surface, is
\begin{eqnarray}
\Omega_C = \int dt \left[ \pm \frac{\pi_{\phi t}}{\sqrt{\pi_{\phi
t}^2 + U(\lambda_t)}} \delta \pi_{\phi t}\wedge \delta \l_t
+\delta \pi_{\phi t} \wedge \delta \phi_t \right].
\end{eqnarray}

For simplicity of the expressions we introduce the coordinate
\begin{eqnarray}
f_{\pm t} := \phi_t \pm \pi_{\phi t} \int_{0}^{\l_t}
\frac{dx}{\sqrt{\pi_{\phi t}^2 + U(x)}}, \label{f+-}
\end{eqnarray}
 and we then write the expression for the restricted symplectic
 form as
\begin{eqnarray}
\Omega_C = \int dt \delta \pi_{\phi t} \wedge \delta f_{\pm t}.
\end{eqnarray}
The vector fields $\frac{\delta}{\delta \pi_{\phi t}}$ and
$\frac{\delta}{\delta f_{\pm t}}$  correspond to non-degenerate
directions. The coordinates $\pi_{\phi t}$ and $f_{\pm t}$ commute
with the constraints, hence they can be projected into functions
$\tilde{\pi}{\phi t}, \tilde{f}_{\pm t}$ on the histories reduced
phase space. However the reduced phase space consists of two
disconnected components, each of topology ${\bf R}^2$. We then
write the symplectic form of the reduced state space as

\begin{eqnarray}
\Omega_{red} = \int dt \delta \tilde{\pi}_{\phi t} \wedge \delta
\tilde{f}_{\pm t}. \label{omegared}
\end{eqnarray}

In order to study the physical interpretation of each component we
consider histories $(\pi_{\phi t}, f_{\pm t}) = (c_1, c_2)$, for
constants $c_1, c_2$ that are solutions to the equations of motion
as we shall see in what follows. These histories correspond to
paths on the constraint surface that satisfy
\begin{eqnarray}
\pi_{\phi t} &=& c_1 \\
\dot{\phi}_t &=& c_2 \mp \frac{c_1}{\sqrt{c_1^2 +U(\l)}} \dot{\l}.
\end{eqnarray}

In order to compare these solutions with known solutions of
Einstein equations we choose a time variable that plays the role
of a clock. This may be achieved by `fixing the gauge' on the
constraint surface \cite{I92, Kuc91}; in this case the causal
ordering parameter $t$ assumes the role of a clock.

The most convenient gauge-fixing condition is to assume that
$\dot{\phi} = \pi_{\phi} e^{-3 \l}$. The parameter $t$ is then
identified to be the global time of Robertson-Walker universes. It
is easy to show that
\begin{eqnarray}
\mp t = \int^{\l}_{\l_0} \frac{e^{3x} dx}{\sqrt{c_1^2 +U(x)}},
\end{eqnarray}
which describes a contracting universe (with respect to $t$) in
the plus sign branch, and an expanding universe in the minus sign
branch. To check this result one may compute the integral for
large $\l$, in which case $e^{\l} \sim e^{\mp \sqrt{\Lambda}t}$.

Finally, note that one may pass from the $+$ component of the
reduced state space to the $-$ component, through the symplectic
transformation $(\pi_{\phi t}, f_{\pm t}) \rightarrow (-\pi_{\phi
t}, f_{\mp t})$.

\subsubsection{$\Lambda > 0$, $\kappa = 1$}

In this case the potential changes sign at $e^{2 \lambda} =
\frac{1}{\Lambda}$, so that $U(\l) > 0$ for $\l > \l^c =
\frac{1}{2} \log(1/\Lambda)$ and $U(\l < 0)$ for $\l < \l^c$. The
potential has a minimum $U_{min} = -\frac{4}{27 \Lambda^2}$ at
$e^{\l_{min}} = \sqrt{\frac{2}{3\Lambda}}$.

We follow the same procedure as in the previous case, however in
this case the constraint surface has a boundary, determined by the
condition
\begin{eqnarray}
\pi_{\l t}^2 = \pi_{\phi t}^2 + U(\l_t) \geq 0.
\end{eqnarray}
The boundary condition is trivially satisfied for $\l > \l^c$ but
it places restrictions on the values of $\pi_{\phi t}$, for $\l <
\l^c$.

The  orbits generated by the constraints have different behavior
in different regions of the constraint surface $C_h$; this
reflects into the non-trivial global structure of the reduced
phase space.

\begin{enumerate}
\item[i.] The regions $U_+$ and $U_-$ correspond to the values of
 $ \pi_{\phi t}^2
> |U_{min}|, \forall t$, and $\pi_{\l}$ being positive and
negative respectively. As in the previous case $U_-$ describes
eternally expanding universes and $U_+$ collapsing ones. This is
due to the fact that for $ \pi_{\phi t}^2
> |U_{min}|$, the orbits of the constraints are
curves with $\l \in ( - \infty, \infty)$. The quantities
$\pi_{\phi t}$ and $f_{\pm t}$ are constant along the orbits and
they project therefore to coordinates on the reduced histories
state space. The symplectic form in this case is the same as Eq.\
(\ref{omegared}).

\item[ii.] For $\pi_{\phi t}^2 < |U_{min}|$ there exist again two
disjoint regions. They both correspond to bouncing universes,
because $\pi_{\l}$ can take both positive and negative values in
the corresponding orbits.  There exist two disjoint orbits for
each value of $\pi_{\phi}$ there exist two disjoint orbits: one
with $\l \in ( - \infty, \l_1(\pi_{\phi})]$ and one with $\l \in
[\l_2(\pi_{\phi}), \infty)$, where $\l_1(\pi_{\phi})$ and
$\l_2(\pi_{\phi})$ the smaller and the larger real solutions of
the equation $\pi_{\phi^2} + U(\l) = 0$. The orbits of the first
type (elements of $U_<$) correspond to an expanding universe that
reaches the critical value $\l_1(\pi_{\phi}$ and re-collapses,
while the orbits of the second type (elements of $U_>$) correspond
to collapsing universes that reach $\l_2(\pi_{\phi}$ and then
re-expands. Note, however, that for $\pi_{\phi} = 0$ there exists
only one solution, which lies in $U_>$. The coordinates $f_+$ and
$f_-$ can be used in both $U_+$ and $U_-$, but are not
independent. In $U_<$ we have
\begin{eqnarray}
f_{ t-} = f_{+ t} + \pi_{\phi t} \int_{-\infty}^{\l_1(\pi_{\phi
t})} \frac{dx}{\sqrt{\pi_{\phi t}^2 + U(x)}},
 \end{eqnarray}
 while
in $U_>$,
\begin{eqnarray}
f_{ t-} = f_{+ t} + \pi_{\phi t} \int_{\l_2(\pi_{\phi
t})}^{\infty} \frac{dx}{\sqrt{\pi_{\phi t}^2 + U(x)}}.
\end{eqnarray}
 The symplectic form is locally the same as in equation Eq.\
(\ref{omegared}).

\item[iii.] There exist two bifurcate surfaces in $\Pi_{red}$, which forms
the boundary between the regions $U_{\pm}$ and $U_<, U_>$. This
boundary is characterised by values of $\pi_{\phi} = \pm \sqrt{
|U_{min}|}$. For each sign of $\pi_{\phi}$ and value of $f_{\pm}$
there exist four different orbits in that surface. Two correspond
to $\l \in (- \infty, \l^{cr})$, one expanding ($\pi_{\l} < 0$)
and one collapsing ($\pi_{\l} > 0$). We will denote them $B_{<-}$
and $B_{<+}$ respectively. The other two correspond to $\l \in
(\l^{cr}, \infty)$ (also collapsing and expanding)--we shall
denote them as $B_{>-}$ and $B_{>+}$.

\item[iv.] Finally there exist two degenerate orbits for $\pi_{\phi} =
\pm \sqrt{ |U_{min}|}, \pi_{\l} = 0$ and for every value of $\phi
= f_{\pm}$. These orbits  only contain the point  $\l = \l^{cr}$,
and  correspond to the static Einstein universe. In this case the
symplectic form diverges; indeed the inclusion of  those orbits in
the reduced state space $\Pi_{red}$ is incompatible with a smooth
manifold structure, and renders $\Pi_{red}$ into an orbifold.
 We shall
denote the corresponding subset of the reduced state space as $O$.

\end{enumerate}

%\begin{tabular}{|c|c|c|}
  % after \\: \hline or \cline{col1-col2} \cline{col3-col4} ...
%  {\bf Region}& {\bf Definition } & {\bf Description} \\
%  & & \\
%  $U_+$ &  $\pi_{\phi }^2 > |U_{min}|, \pi_{\l} > 0 $ & collapsing universe \\
 % $U_-$ & $\pi_{\phi }^2 > |U_{min}|, \pi_{\l} < 0 $ & expanding universe \\
 % $U_<$ & $\pi_{\phi }^2 < |U_{min}|, \l \leq \l_1(\pi_{\phi})$ & bounded bouncing universe \\
 % $U_>$ & $\pi_{\phi }^2 < |U_{min}|, \l \geq \l_2(\pi_{\phi})$ & unbounded bouncing universe \\
 % $B_{<-}$ & $\pi_{\phi }^2 = |U_{min}|, \l < \l^{cr}, \pi_{\l}< 0$ & bounded expanding universe \\
 % $B_{<+}$ & $\pi_{\phi }^2 = |U_{min}|, \l < \l^{cr}, \pi_{\l}> 0$ & bounded collapsing universe \\
 % $B_{>-}$ & $\pi_{\phi }^2 = |U_{min}|, \l > \l^{cr}, \pi_{\l}< 0$ & unbounded expanding universe \\
 % $B_{>+}$ & $\pi_{\phi }^2 = |U_{min}|, \l > \l^{cr}, \pi_{\l}> 0$ & unbounded collapsing universe \\
 % $O$ &   $\pi_{\phi }^2 = |U_{min}|, \l = \l^{cr}$ & Einstein static universe \\ \hline
%\end{tabular}

\subsection{Equations of motion and time-reparameterisation}

It is easy to verify that the action functional $S$ commutes
weakly with the constraints, and thus projects to a function
$\tilde{S}$ on the histories reduced state space $\Pi_{red}$. In
fact $\tilde{S} = \tilde{V}$, where $\tilde{V}$ is the projection
of $V$ onto $\Pi_{red}$. The projected Liouville function
$\tilde{V}$ still generates time translations on the reduced state
space. Indeed,
\begin{eqnarray}
\tilde{S} = \tilde{V} = \int dt \pi_{\phi t} \dot{f}_{\pm t}.
\end{eqnarray}

The existence of a symplectic representation of time-translations
on $\Pi_{red}$ highlights one of the most important properties of
the histories formalism: time-ordering is not lost on the space of
true degrees of freedom. Indeed, the parameter $t$ that determines
time ordering on $\Pi$ is the same with the parameter $t$ that
determines time-ordering on $\Pi_{red}$.

The equations of motion Eq.\ (\ref{eqnsofmotn}) on the reduced
state space have the solutions $\pi_{\phi t} = const.$ and $f_{\pm
t} = const$.

Since the classical paths correspond to constant values of their
parameters, the solutions to the equations of motion are invariant
under time reparameterisations. The issue arises, whether there
exists a symplectic action of the group of
time-reparameterisations that can be projected on the histories
reduced state space.

To this end, the natural object to study is the family of
functions $V_U$
\begin{eqnarray}
V_U = \int dt (\pi_{\l t} {\cal L}_U \l_t + \pi_{N t} {\cal L}_U
N_t + \pi_{\phi t} {\cal L}_U \phi_t), \label{failedrepar}
\end{eqnarray}
where $U = U(t) \frac{\partial}{\partial t}$ is a vector field on
$\R$. While these functions satisfy the algebra of the spacetime
diffeomorphisms group ${\rm Diff}(\R)$, they do not commute weakly
with the constraints, hence they cannot be projected on to the
reduced state space. The reason is that they fail to reproduce the
correct behaviour on $\sqrt{h_t}$, namely
\begin{eqnarray}
\{V_U, \sqrt{h_t} \} = {\cal L}_U \sqrt{h_t}.
\end{eqnarray}
While $\sqrt{h}$ is a scalar density of weight one, it transforms
under $V_U$ as a scalar function on $\R$, and consequently the
constraints do not transform in a proper geometric way under the
action of $V_U$. The only functions $V_U$ that project into the
reduced state space are those for which the condition $U(t) =
constant$ holds; these functions are multiples of the Liouville
generators of time translations.

Nonetheless a representation of time reparameterisations on the
reduced state space does exist, but its  generators are the
functions $V_W$ of equation Eq.\ (\ref{minidiff}) rather than the
ones of Eq. (\ref{failedrepar}). The functions (\ref{minidiff})
 generate a representation of $Diff(\R)$ on $\Pi_{RW}$. If we assume that
the foliation map ${\cal E}$ depends on the metric and satisfies
the equivariance condition Eq.\ (\ref{Def:EquivariantFoliation2}),
the functions $V_W$ commute with the constraints and can therefore
be projected on $\Pi_{red}$--see \cite{Sav03b}.

We must emphasise that the requirement of a metric-dependent
foliation is only necessary in general relativity; it is
superfluous in the minisuperspace model and it only appears in the
discussion of the invariance of $\Pi_{red}$ under diffeomorphisms.
For this reason any parameterised system that does not possess an
associated covariant diffeomorphic description is not expected to
carry a symplectic action of the group ${\rm Diff}(\R)$ on the
space of the true degrees of freedom $\Pi_{red}$.

\section{Histories quantisation}

In this section we provide a summary of the histories quantisation
procedure, emphasising the methods for constructing the histories
Hilbert space and the decoherence functional. We extend the
discussion to the case of systems characterised by Hamiltonians
without a vacuum state--this is the case relevant to the
quantisation of minisuperspace models. We also elaborate on the
physical distinction between different choices for the history
Hilbert space.

\subsection{The history group}
 The key object in histories quantisation is the histories
Hilbert space ${\cal V}$. In the case of discrete-time
histories---namely for histories with support on the moments of
time $(t_1, t_2, \ldots, t_n)$---the history Hilbert space ${\cal
V}$ is the tensor product of the single-time Hilbert spaces
$H_{t_1} \otimes H_{t_2} \otimes \ldots \otimes H_{t_n}$, where
$H_{t_i}$ is a copy of the canonical Hilbert space of the theory,
at time $t_i$. We recall that a history is represented by a
projection operator on ${\cal V}$.

In the case of gravity  we consider a continuous temporal support
for the histories; the full real line $\R$ or perhaps an interval
$[0,T]\in \R$. This suggests defining a continuous-tensor product
of Hilbert spaces, which is a rather unwieldy object, being in
general non-separable.

Isham and Linden \cite{IL95} proposed a construction that
addresses this issue. They constructed the continuous-time Hilbert
space by seeking the representations of the {\em history group\/},
a history analogue of the canonical group. For a particle at a
line\cite{IL95}. the Lie algebra of the history group is
\begin{eqnarray}
 [\hat{x}_t, \hat{x}_{t'} ] &=& 0 \\
 \[\hat{p}_{t},\hat{p}_{t'} ] &=& 0 \\
 \[\hat{x}_t, \hat{p}_{t'} ] &=& i \delta (t,t').
\end{eqnarray}

For an harmonic oscillator, one may select uniquely the Hilbert
space if one assumes the existence of an operator representing the
smeared Hamiltonian \cite{ILSS98}. The result depends on important
properties of quadratic Hamiltonians---it can be generalised to
deal with systems that are described canonically by Fock spaces.

A crucial property of these constructions is that the Hilbert
space that carry a representation of the history group can be
casted in a tensor product-like form as ${\cal V} = \otimes_t
H_t$. This is not a genuine tensor product but rather a specific
mathematical object that can be rigorously defined. One may employ
the tensor-product-like structure of the history Hilbert space
 to find more general representations of the history group \cite{An01b}. We will employ
such representations in the study of
 the
minisuperspace models. The construction of the history Hilbert
space relies heavily on the theory of coherent states, therefore
we shall give a brief summary on the theory of coherent states for
the study of Hilbert space representations-- the reader is
referred to references \cite{KlSk85, Per86} for details.

\subsection{Coherent states}
We assume a representation of the canonical or history group, by
unitary operators $\hat{U}(g)$ on a Hilbert space ${\cal H}$.
Furthermore, let $\hat{h}$ denote the Hamiltonian of this system
and by  $|0 \rangle_H$ the vacuum state, i.e. the  Hamiltonian's
lowest eigenstate. Then we define the coherent states as the
vectors
\begin{equation}
|g \rangle := \hat{U}(g) | 0 \rangle .
\end{equation}
Next we consider the equivalence relation on the canonical group
defined as\\

 \hspace*{1cm}$g \sim g'$ \hspace*{0.7cm}if \hspace*{0.7cm}
 $|g \rangle$, $|g' \rangle$ \hspace*{0.1cm} correspond to the
same ray.\\

The phase space $\Gamma$ is identified as the quotient space $G/
\sim $ and we label the (generalised) coherent states by points $z
\in \Gamma$.

The fundamental property of coherent states is that they form an
overcomplete basis, i.e., any vector $|\psi \rangle$ can be
 written as
\begin{equation}
|\psi \rangle = \int d \mu(z) f(z) |z \rangle ,
\end{equation}
in terms of some complex-valued function $f$ on $\Gamma$. Here $d
\mu$ denotes some natural measure on $\Gamma$. There is also a
decomposition of the unity in terms of the coherent states
\begin{equation}
\int d \mu(z) |z \rangle \langle z| = \hat{1}.
\end{equation}

Another important property of coherent states is that the overlap
kernel $ \langle z'| z \rangle$ contains all the information about
the Hilbert space and the group representation.

Let us denote an overlap kernel by $K(z|z')$. This may be a
general function on $\Gamma \times \Gamma$. However, it has to
satisfy two properties in order to correspond to coherent states.
First, it has to be hermitian, $K(z|z') = K^*(z'|z)$. Second, it
needs to be positive-definite, that is, for any sequence of
complex numbers $c_n$ and points $z_n \in \Gamma$,
\begin{eqnarray}
\sum_{nm} c^*_n c_m K(z_n|z_n) \geq 0.
\end{eqnarray}
The last condition implies that $K(z,z) \geq 0$. We can now
construct the corresponding Hilbert space ${\cal H}$ in the
following way.

A vector on $\cal{H}$ can be constructed as a function on phase
space $\Gamma$ of the form $\Psi(z) = \sum_l c_l \langle z_l| z
\rangle $, for a finite number of complex numbers $c_l$ and state
space points  $z_l$. The inner product between two vectors
characterised by $c_l, z_l$ and $c'_l, z'_l$ is
\begin{eqnarray}
 \sum_l c'^*_l c_l \langle z'_l|z_l \rangle .
\end{eqnarray}
The wave functions $\Psi_{z'}(z) = K(z'|z)$ form a family of
coherent states on $\cal{H}$.

 The canonical group is then represented by the operators
$\hat{U}(g)$, which are defined by their action on coherent states
\begin{equation}
(\hat{U}(g) \Psi_{z'})(z) = e^{i \phi} \Psi_{g z'} (z),
\end{equation}
where the precise choice of phase depends on the details of the
group structure. The expression $g z$ denotes the action of the
group element $g$ on the manifold $\Gamma$.

\subsection{Constructing general representations of the history group}
Let us denote the space of temporal supports by $\cal{T}$, which
can be either $\R$ or and interval $[0,T]\in \R$. We construct the
history Hilbert space ${\cal V}$ by selecting a family $L$ of
normalised coherent states $|z \rangle$, on the Hilbert space of
the canonical theory \cite{An01b}.

Next we define the space of all continuous paths from $S$ to $L$.
We write such paths $t \rightarrow |z \rangle_t$ as $|z(.)
\rangle$. We then construct the vector space $V$ of {\em finite }
linear combinations of the form $\sum_{i = 1}^l |z_l(\cdot)
\rangle$, which is equipped with the inner product
\begin{eqnarray}
\langle z(\cdot)| z'(\cdot) \rangle = \exp \left(\int_S dt \log
\langle z_t| z'_t\rangle \right). \label{innerprod}
\end{eqnarray}
In order to obtain finite values for this inner product we need to
place some restrictions on the possible paths $z(\cdot)$, for the
case $ \cal{T} = \R$. In particular we assume that $\langle
z_t|z'_t \rangle$ converges rapidly to 1, for $t \rightarrow \pm
\infty$; or else, all paths converge asymptotically to a fixed
value of $z_0$. When $\cal{T}$ is a bounded interval, there are no
restrictions. We then employ $\langle z(\cdot)| z'(\cdot) \rangle$
as an overlap kernel for the coherent states $|z(\cdot)\rangle$.
Then we use the definition of the overlap kernel for the
construction of the history Hilbert space and the corresponding
representation of the history group.

We define the Hilbert space ${\cal V}$ by taking the closure of
$V$ with respect to the topology induced by the inner product Eq.\
(\ref{innerprod}). If the canonical coherent states are obtained
by the action of a representation of the canonical group $G$, then
the Hilbert space ${\cal V}$ carries a representation of the
corresponding history group.

For example, let us consider a generator $\hat{A}$ of the
canonical group on $\cal{H}$. Then we may define the generator
$\hat{A}_f = \int dt \hat{A}_t f(t)$ of the corresponding history
group, which is smeared with respect to a function $f(t)$. To this
end, we write the matrix elements of the corresponding
one-parameter group of automorphisms as
\begin{eqnarray}
\langle  z(\cdot)|e^{i\hat{A}_f s} |z'(\cdot) \rangle = \exp
\left( \int dt  \log \langle z_t|e^{i\hat{A}f(t) s}| z'_t \rangle
\right). \label{historyoperator}
\end{eqnarray}
It is possible to show that $e^{i\hat{A}_f s}$ is continuous at $s
= 0$ for a large class of functions $f$\cite{An01b}. Hence,
according to Stone's theorem, the operator ${\cal V}$ exists as a
self-adjoint operator on ${\cal V}$.

\bigskip

The role of coherent states is crucial in this construction. We
could have chosen to construct a representation of the Hilbert
space by introducing another arbitrary family $L$ of vectors on
the Hilbert space, and then repeat the same algorithm. However if
the basis formed by this set of vectors was too small, then the
resulting history space would not be able, in general, to carry
many interesting self-adjoint history operators,---like the
generators of the history group. On the other hand, if the set $L$
was taken to be too large,(say the whole Hilbert space $H$), the
resulting Hilbert space would again be too large and many
interesting {\em unbounded} self-adjoint operators---such as the
smeared Hamiltonian---could not have been defined. The choice of
the set of vectors $L$ to be a family of {\em coherent states\/}
seems to be of exactly the right size.

However, different sets of coherent states may lead to different
history Hilbert spaces and different representations of the
history group. We must choose the set of coherent states,
therefore, by taking into account the symmetries of the canonical
theory, and in some cases---for example the presence of a
Poincar\'e group symmetry and a Poincar\'e invariant vacuum---this
choice might be rendered unique \cite{Sav01b}.

In the minisuperspace approximation, however, most traces of the
original symmetry of general relativity have been lost. As a
result the representation of the history group is non-unique.

\subsection{Defining history operators}
It is straightforward to define the analogue of the histories
time-translation generator $V$ on the history Hilbert space ${\cal
V}$. $V$ is defined through the one-parameter group of unitary
operators
\begin{eqnarray}
e^{is \hat{V}} | z (\cdot) \rangle = | z'(\cdot) \rangle,
\end{eqnarray}
where $z'(t) = z(t+s)$. This definition is possible when the space
of temporal supports $\cal{T}$ is equal to the whole real line.

The definition of the time-averaged Hamiltonian `restricts' the
arbitrariness in the choice of the history group representation,
even to the extent of leading to a unique representation. Indeed,
if the Hamiltonian $\hat{h}$ of the canonical theory possesses a
unique vacuum state $| 0 \rangle$ such that $\hat{h}|0 \rangle =
0$, we may define the coherent states by taking $|0 \rangle$ as a
reference vector. We may then define the one-parameter group of
transformations $e^{i \hat{H}_{\kappa}}$ by using the matrix
elements
\begin{eqnarray}
\langle z(\cdot)|e^{-i \hat{H}_{\kappa}s}|z'(\cdot) \rangle = \exp
\left( \int_S dt \log \langle z_t|e^{-i\hat{h} \kappa(t)s}|z'(t)
\rangle \right). \label{Hammatel}
\end{eqnarray}
We can easily prove that the  define a self adjoint operator if
$\cal{T}$ is a finite interval. Alternatively, if $\cal{T} = \R$,
$\hat{H}_{\kappa}$ can be defined  if $|z_t\rangle$ converge
asymptotically ($t \rightarrow \pm \infty$) to $| 0 \rangle$.

If the canonical vacuum is invariant under a symmetry of space
translations then the definition of $H_{\kappa}$ unique selects
the representation of the history group\cite{An01b}.

In the case of the minisuperspace model the Hamiltonian operator
does not have a continuous spectrum around zero. Hence we cannot
define a vacuum state that could be used to select a preferred
family of coherent states. The family of coherent states has to be
chosen with reference to other symmetries of the theory.

In absence of a unique choice of coherent states we may still
employ the Eq.\ (\ref{Hammatel}) for defining the smeared
Hamiltonian $H_{\kappa}$. It is easy to show that
\begin{eqnarray}
|\langle z(\cdot)|e^{-i \hat{H}_{\kappa}s} - 1 |z(\cdot) \rangle|
< C s \int_S dt \kappa(t) |\log \langle z_t|\hat{h} |z_t \rangle|,
\label{integral}
\end{eqnarray}
for some constant $C > 0$ for sufficiently small values of $s$.

If $\cal{T}$ is a finite interval, or if the smearing function
$\kappa(t)$ has compact support then the integral Eq.\
(\ref{integral}) takes a finite value. Hence the one-parameter
group of transformations is continuous at $s = 0$, and according
to Stone's theorem, the operator $\hat{H}_{\kappa}$ exists.

If $\cal{T} = \R$ and if the time-averaging function $\kappa(t)$
does not vanish fast enough at infinity, then the integral Eq.\
(\ref{integral}) diverges, since the quantity  $\langle
z_t|\hat{h} |z_t \rangle$ tends to the non-zero value $\langle
0|\hat{h}|0 \rangle$ at infinity. Hence we cannot prove the
existence of the $\hat{H_{\kappa}}$ operator, while if $\hat{h}$
possessed $|0 \rangle$ as an eigenstate $\hat{H}_{\kappa}$ would
be definable for any measurable function $\kappa(t)$\footnote{We
can always redefine $\hat{h}$ up to an additive constant so that
$\langle 0| \hat{h} |0 \rangle = 0$, but this does not remove the
ambiguity in the choice of a family of coherent states. Such a
redefinition is also  inadequate for the case that $\hat{h}$
represents a constraint, because the addition of a constant to a
constraint produces a constraint with a different physical
content.}.

Once we have defined the self-adjoint operators $\hat{V}$ and
$\hat{H}_{\kappa}$ we can easily show that they satisfy the
commutation relations
\begin{eqnarray}
[\hat{V}, \hat{H}_{\kappa} ]  = i \hat{H}_{\dot{\kappa}}.
\end{eqnarray}

\subsection{The decoherence functional}
In the histories formalism the probabilities are contained in the
decoherence functional. The decoherence functional incorporates
the information about the initial state, the dynamics and the
instantaneous laws. We expect therefore to implement the
constraints in the construction of the decoherence functional.

The decoherence functional is a complex-valued function of a pair
of histories. A history may be represented by a projection
operator $P$ on ${\cal V}$, so the decoherence functional is a
function of a pair of projection operators. A projector may be
written in terms of fine-grained, one-dimensional projectors of
the form $|f \rangle \langle f|$, by means of coarse-graining
operations. Furthermore, any vector $| f \rangle \in {\cal V}$ may
be written as a linear combination of coherent states. For this
reason, it is sufficient to compute the decoherence functional for
a pair of projectors onto coherent states.

There exists a rather general  theorem about the form of the
decoherence functional in a physical system \cite{Ana03}. If the
decoherence functional satisfies a version of the Markov
property--namely that expectations of physical observables at a
moment of time allow the determination of expactations at all
future moments of time-- then it can be obtained  as a suitable
continuous limit of the following expression

\begin{eqnarray}
d(z_1,t_1; z_2,t_2; \ldots ; z_n, t_n|z'_1,t'_1;
z'_2, t'_2; \ldots ; z'_m,t'_m) =  \nonumber \\
\langle z'_m|e^{-i \hat{H}(t'_m - t_n)}| z_n \rangle \langle z_n|
e^{-i \hat{H}(t_{n}-t_{n-1})}| z_{n-1} \rangle \ldots
\nonumber \\
\times \langle z_2|e^{-i \hat{H}(t_2-t_1)}| z_1 \rangle \langle
z_1|e^{-i \hat{H}(t_1-t_0)}| z_0 \rangle \langle
z_0|\hat{\rho}_0|z'_0
\rangle\nonumber \\
\times \langle z'_0| e^{i \hat{H}(t'_1-t_0)}| z'_1 \rangle \langle
z'_1|e^{i\hat{H}(t'_2-t'_1)} | z'_2 \rangle \ldots \langle
z'_{n-1}|e^{i \hat{H}(t'_n-t'_{n-1})}| z'_n \rangle.
\end{eqnarray}
The decoherence functional takes value in an $n-point$ and an
$m-point$ history, while $\hat{\rho}_0$  is the initial state.

In the continuous-time limit ($|t_i - t_{i-1}| < \delta t$, for
all $i$ and $\delta t \rightarrow 0$) the decoherence functional
becomes
\begin{eqnarray}
d[z(\cdot), z'(\cdot)] \sim e^{iS[z(\cdot)] - iS[z'(\cdot)]},
\end{eqnarray}
where  $S = V - \int dt h(z)$ is the classical action functional
and $h(z,z) = \langle z|\hat{h}|z \rangle$.

For differentiable paths $z(\cdot)$ the choice of the family of
coherent states (and hence the representation of the history
group) does not affect the values of the decoherence functional.
The difference between representations appears for
non-differentiable paths. In this case, one has to keep terms of
order $\delta t^2$; then it can be shown that different
representations provide different contributions\cite{AnSav03}.

The last result is particularly relevant in the study of
coarse-grained histories. In order to construct a coarse-grained
history we have to sum over coherent state paths. This involves
employing an integration measure which, by necessity, has support
on non-differentiable paths (cylinder sets), and provides
different results for different families of coherent states.

In conclusion, different choices of coherent states families,
implies different representations for the history group. All
representations---with the same Hamiltonian---provide the same
values for the decoherence functional for fine-grained,
differentiable coherent state histories. However they involve
different rules for coarse-graining and hence provide different
answers to probabilities for coarse-grained histories. It is very
important that physical criteria should exist, that allow the
unique selection of a representation for the history group.

\section{The implementation of constraints}

In order to study the quantisation of parameterised systems such
as the minisuperspace model, one has to extend the previous
analysis to deal with systems that possess first class
constraints. In section 4.5 we noted that if the quantum history
theory description of a system is assumed to satisfy a version of
the Markov property, then the quantum description of the system is
obtained from the knowledge of the coherent states of the standard
theory . The ambiguity in the choice of a family of coherent
states can be in principle resolved---as it does in a large class
of systems that have been studied\cite{Sav01b, An01b, Bur04}---by
the knowledge of the symmetries of the fully covariant theory.

In the case of gravity this would imply that we possess the
knowledge of the full quantum gravity theory---at least at the
kinematical level. In particular, it would mean that we have
explicitly constructed coherent states for the history variables
of general relativity $|g_{\mu\nu}(\cdot),\pi^{\mu \nu}(\cdot)
\rangle$, in such a way as to implement the principle of general
covariance---the existence of a unitary representation of the
$Diff(M)$ group whose generators commutes with the history version
of the canonical constraints---as we have in classical history
theory. We would then restrict the definition of these coherent
states for the special case of a R-W metric.

In absence of a theory for the full spacetime symmetries of
general relativity, the quantisation of minisuperspace models is
beset with a degree of arbitrariness, which is related to the
choice of a proper integration measure of coarse-grainings. Next
we analyse this issue in detail.

There exist two main general schemes for the quantisation of
constrained systems that do not involve gauge fixing: Dirac
quantisation and reduced state space quantisation. The former
implements the constraints at the quantum mechanical level, while
the latter implements them classically and then attempts to
quantise the classical reduced state space. We study both schemes
in the following two subsections.

\subsection{Dirac quantisation}
In \cite{SavAn00} we analysed the transcription of the Dirac
quantisation method in the history context. The general idea is to
first construct the Hilbert space of the unconstrained system, and
then to identify an operator that represents the canonical
constraint smeared in time; the zero-eigenspace of the constraint
operator is the physical Hilbert space.

Specifically, if $\cal{H}$ is the single time Hilbert space of the
standard canonical theory, equipped with a set of coherent states
$| z(\cdot) \rangle$, and if ${\bf E}$ is the projector into the
zero eigenspace of the constraint, then, the matrix elements of
${\bf E}$ of the coherent states $\langle z|{\bf E}|z' \rangle$
define a new overlap kernel; from this kernel one may define the
physical Hilbert space \cite{Kla97, KeKl01}.

The Hilbert space ${\cal V}$ of the corresponding history theory
is then spanned by coherent state paths $|z(\cdot) \rangle$, which
may be employed to define the projector ${\bf E}_{his}$ to the
history physical Hilbert space
\begin{eqnarray}
\langle z(\cdot) | {\bf E}_{his}| z'(\cdot) \rangle = \exp \left(
\int_S dt \log \langle z_t| {\bf E_t}|z'_t \rangle \right).
\end{eqnarray}

Then we define the decoherence functional for coherent state paths
as
\begin{eqnarray}
d(z_1,t_1; z_2,t_2; \ldots ; z_n, t_n|z'_1,t'_1;
z'_2, t'_2; \ldots ; z'_m,t'_m) =  \nonumber \\
\langle z'_m|{\bf E}| z_n \rangle \langle z_n| {\bf E}| z_{n-1}
\rangle \ldots \langle z_2|{\bf E}| z_1 \rangle \langle z_1|{\bf
E}| z_0 \rangle \langle z_0|{\bf E}\hat{\rho}_0 {\bf E}|z'_0
\rangle  \nonumber \\
\langle z'_0| {\bf E}| z'_1 \rangle \langle z'_1| {\bf E} | z'_2
\rangle \ldots \langle z'_{n-1}| {\bf E}| z'_n \rangle.
\end{eqnarray}

Writing formally ${\bf E} = \frac{1}{2 \pi}\int d\xi e^{i \xi
\hat{h}}$ and inserting this expression at every time step  we may
write the decoherence functional in the continuous limit as
\begin{eqnarray}
d[z(\cdot), z'(\cdot)] = \langle z_0|\hat{\rho}_0| z_0 \rangle
\int D \xi(\cdot) D \xi'(\cdot) e^{i (V - H_{\xi})[z(\cdot)]  - i
(V - H_{\xi'})[z'(\cdot)]}, \label{decfundirac}
\end{eqnarray}
where $D \xi(\cdot) $ is the continuous limit of $\frac{d
\xi_{t_1} d \xi_{t_2} \ldots d \xi_{t_n}}{(2 \pi)^n}$. This
expression is equivalent to the decoherence functional of a system
with smeared Hamiltonian $H_{\xi}$, integrating however over all
possible paths of the Lagrange multiplier $\xi$.

A weak point of the Dirac quantisation is that, typically the
canonical constraint possesses continuous spectrum near zero.
Hence, the physical Hilbert space is not a subspace of the initial
Hilbert space. To deal with such problems one has to employ more
elaborate techniques, usually involving the concept of the induced
inner product,---see \cite{HaMa97} for a discussion close to our
present context.

The kernel $K(z,z')$ has degenerate directions, namely there exist
points $z_1$ and $z_2$, such that $|K(z_1,z_2)|^2 = 1$. This
implies that we have to quotient the parameters of the state space
with respect to the equivalence relation\\

\hspace*{3cm}$z \sim z'$ \hspace*{0.5cm}if
\hspace*{0.5cm}$|K(z,z')|^2 = 1$.\\

For a large class of constraints this results to an overlap kernel
defined over the reduced state space\cite{Kla97}; this property is
not in general guaranteed.

In any case, we obtain a family of coherent state paths
$|\zeta(\cdot) \rangle$ on $V_{phys}$, and a map from $V$ to
$V_{phys}$ implemented through the coherent states as
$|z(\cdot)\rangle \rightarrow |\zeta(\cdot) \rangle $, where
$\zeta$ corresponds to the equivalence class in which $z$ belongs.

We exploited the Dirac method, in \cite{SavAn00}, for the simple
example of the relativistic particle. However, we have found a
more convenient quantisation method, that entails ideas from both
the reduced state space and the Dirac quantisation and it seems to
be more suitably adapted to the needs of the histories scheme.

\subsection{Reduced state space quantisation}
The key idea of the reduced state space quantisation is to
implement the constraints {\em before} quantisation, at the level
of the classical state space, and then to seek an appropriate
quantum representation for the reduced state space.

It is easier, in general, to solve the constraints in classical
theory than in the quantum theory---we have for example no
problems with the continuous spectrum of operators or issues
related to normal ordering. Nevertheless, the reduced state space
is usually a manifold of non-trivial topological structure (not a
cotangent bundle); often it is not even a manifold, if it
possesses orbitfold-like singularities.

The reduced state space quantisation scheme has also other
drawbacks. For example, the reduced state space of field systems
generically consists of non-local variables: the true degrees of
freedom are not fields themselves. For this reason the spacetime
character of the theory is not explicitly manifest in the reduced
state space. We have then no way to represent fields quantum
mechanically and to study their corresponding symmetries.

Histories general relativity is an exception to this rule though,
because the spacetime diffeomorphisms group is represented both on
the unconstrained and on the reduced state space\cite{Sav03b}.

Since histories quantisation---in the form we presented
here---relies strongly on the coherent states, it may be possible
to exploit the fact that they provide a link between the classical
state space description and the quantum mechanical one on a
Hilbert space. Then we may implement the constraints at the
classical level, while wpreserving the basic structures of Dirac
quantisation, namely the coexistence of a Hilbert
space---corresponding to the full classical phase space---with the
Hilbert space of the true degrees of freedom.

The resulting construction is a hybrid between the reduced state
space quantisation and Dirac quantisation. It will be presented in
full details elsewhere\cite{Ana04}. Here we shall explain the
method with reference to the histories construction.

For the construction of a histories Hilbert space, of the true
degrees of freedom, it is sufficient to identify a set of coherent
states on the reduced state space of the single-time theory. This
point was fully explained in the previous section.

Let us denote as $|\zeta \rangle$ these coherent states, $\zeta
\in \Gamma_{red}$. All information about the quantum theory is
contained in the overlap kernel $K_{red}(\zeta,\zeta')$ which is a
function on $\Gamma_{red}\times \Gamma_{red}$.

In Dirac quantisation we start with the full Hilbert space, which
contains a family of coherent states $|z \rangle $. The associated
overlap kernel $K(z,z')$ is a function on $\Gamma \times \Gamma$,
where $\Gamma$ is the state space of the system before the
imposition of the constraints. The overlap kernel contains all
information about the Hilbert space of the system and of the
physical obsrvables. The important issue is, then, to find a
procedure in order to pass from a function on $\Gamma \times
\Gamma$ to a function of $\Gamma_{red} \times \Gamma_{red}$, in
accord with the geometric definition of the reduced state space.

We shall employ here a specific procedure that is based upon the
geometric structure induced by the coherent states on the
classical state space. Of relevance is the limiting behaviour of
the overlap kernel $\langle z|z' \rangle$ when $z' = z +\delta z$,
\begin{equation}
\langle z| z + \delta z \rangle = \exp \left( i A_i(z +
\frac{\delta z}{2}) \delta z^i - \frac{1}{2} g_{ij} \delta z^i
\delta z^i \right) + O(\delta z^3),
\end{equation}
and where $A_i$ is a U(1)-connection one-form  and $g_{ij}$ is a
Riemannian metric on the classical state space
\begin{eqnarray}
i A_i(z) &=& \langle z|\partial_i z \rangle, \\
g_{ij}(z) &=& \langle \partial_i z| \partial_j z\rangle + A_i(z)
A_j(z).
\end{eqnarray}
The important point here is that this short-distance behavior
allows one to fully reconstruct the overlap kernel as a
path-integral \cite{KlDa84, Kla88}
\begin{eqnarray}
\langle z|z' \rangle = \lim_{\nu \rightarrow \infty} {\cal
N}_{\nu}(t) \int Dz(\cdot)  e^{i \int A  - \frac{1}{2 \nu}
\int_0^t ds g_{ij} \dot{z}^i \dot{z}^j }, \label{pathintegral}
\end{eqnarray}
where ${\cal N}_{\nu}(t)$ is a factor entering for the purpose of
correct normalisation. In other words, the metric on phase space
defines a Wiener process on phase space, which may be employed to
regularise the usual expression for the
coherent-state-path-integral. An overlap kernel then needs two
inputs for its construction: the connection one-form on state
space and the Riemannian metric.

In the context of the reduced quantisation procedure, the one-form
$A$ may be easily defined on the reduced state space, because it
is a symplectic potential of the corresponding symplectic form.

The key point is to identify a metric $\tilde{g}$ on the reduced
state space, starting from a metric $\g$ on the unconstrained
state space $\Gamma$. The key observation is that an element of
the reduced state space---being an orbit of the constraint's
action---is a submanifold of $\Gamma$. Two such submanifolds do
not intersect, because different equivalence classes are always
disjoint. One may then define a distance function between any two
such submanifolds
\begin{eqnarray}
D_{red}(\zeta, \zeta') = \inf_{z \in \zeta, z' \in \zeta'}
D(z,z'),
\end{eqnarray}
where $\zeta, \zeta' \in \Gamma_{red}$ and $D$ is the distance
function on $\Gamma$ corresponding to the metric $ds^2$ on
$\Gamma$. The distance function $D_{red}$ on $\Gamma_{red}$
defines a metric on $\Gamma_{red}$ that may be employed in the
construction of the corresponding overlap kernel via the
path-integral Eq.\ (\ref{pathintegral}).

\subsection{The R-W minisuperspace model}
Next we shall apply the scheme sketched above\cite{Ana04} to the
R-W minisuperspace model.

We first select the family of coherent states on the single-time
Hilbert space  $H$. We have previously explained that this choice
should reflect the symmetries of the underlying theory. To the
extent that the spacetime diffeomorphism symmetry of gravity has
been lost in the reduction to the homogeneous and isotropic model
we consider here, the choice is more or less arbitrary. We find
convenient to employ the standard Gaussian coherent states,
parameterised by $\l,\pi_{\l}, \phi, \pi_{\phi}, N, \pi_N$ with an
overlap kernel
\begin{eqnarray}
&& \langle \l,\pi_{\l}, \phi, \pi_{\phi}, N, \pi_N|\l',\pi_{\l}',
\phi', \pi_{\phi}', N', \pi_{N}' \rangle = \nonumber \\
&& \exp \left( i \pi_{\phi} \phi' - i \pi_{\phi}' \phi + i
\pi_{\l} \l' - i \pi_{\l}' \l  +i \pi_N N' - i \pi_N' N \right.
\\ \nonumber
 &&\left. -\frac{1}{2}[ (\l \!- \!\l')^2 \!+ \!(\pi_{\l}\! -\!
 \pi_{\l}')^2 \!+ \!(\phi \!-\! \phi')^2 \!- \!(\pi_{\phi} \!- \!
 \pi_{\phi}')^2] \!- \!(N \!-\! N')^2 \!-\!(\pi_N \!-\!\pi_N')^2
 \right) \label{coherent}
\end{eqnarray}

The corresponding state space metric is
\begin{eqnarray}
ds^2 = \frac{1}{2} (d \l^2 + d \pi_{\l}^2 + d \phi^2 + d
\pi_{\phi}^2 + dN^2 + d\pi_N^2).
\end{eqnarray}
The elements of the reduced state space correspond to surfaces of
fixed $\pi_{\phi}, f_{\pm}$, while $\pi_{\l} = \pm
\sqrt{\pi_{\phi}^2 + U(\l)}$. Each point of such a surface is
characterised by the values of the parameters $\l$ and $N$. The
$N$ dependence is trivial and we shall disregard it. Hence the
distance between two surfaces characterised by $f_{\pm},
\pi_{\phi}$ and $ f'_{\pm}, \pi_{\phi}'$ respectively, is
\begin{eqnarray}
&& D(f_{\pm}, \pi_{\phi}; f'_{\pm}, \pi_{\phi}') = \inf_{\l, \l'}
\frac{1}{2}[ (\l\! - \!\l')^2 + (\!\sqrt{\pi_{\phi}^2 + U\!(\l)} -
\sqrt{\pi_{\phi}'^2 + U\!(\l')})^2 \; \; \; \; \; \; \; \; \;\;\;\;\;\\
\nonumber
 &&+ (f_{\pm} -
f_{\pm}' \mp \pi_{\phi} \!\int^{\l}\!\!\!\!\!
\frac{dx}{\sqrt{\pi_{\phi}^2 + U\!(x)}} \pm \pi_{\phi}'\!
\int^{\l'}\!\!\!\!\! \frac{dx}{\sqrt{\pi_{\phi}'^2 + U\!(x)}})^2-
(\pi_{\phi}\! -\! \pi_{\phi}')^2]. \label{inf}
\end{eqnarray}

To obtain the metric $d\tilde{s}^2$ on $\Gamma_{red}$, we need to
take the infimum in the above expression for neighbouring orbits,
namely the ones for which $f_{\pm}' = f_{\pm} + \delta f_{\pm}$
and $\pi_{\phi}' = \pi_{\phi} + \delta \pi_{\phi}$.

First we consider the case that the parameter $\l$ takes value in
the full real axis in every orbit. This is the case for $\kappa =
0 , -1$, but also for $\kappa = +1$ in the regions $U_+$ and $U_-$
of the reduced state space.

In this case we observe that, neighbouring orbits  converge as $\l
\rightarrow \infty$, while they diverge exponentially at $\l
\rightarrow - \infty$. It is easy to show that the infimum is
achieved for $\l' \rightarrow \l$ as $\l \rightarrow \infty$. This
leads to the metric
\begin{eqnarray}
d\tilde{s}^2 = \frac{1}{2}(d\tilde{\pi}_{\phi}^2 + d
\tilde{f}_{\pm}^2). \label{redmetric}
\end{eqnarray}

For the $U_<$ and $U_>$ regions of the reduced state space in the
case $\kappa = +1$, we should recall that the parameter $\lambda$
takes values in $(-\infty, \l_1]$ and $[\l_2, \infty)$
respectively. The corresponding orbits tend to converge near
$\l_1$ and $\l_2$ and the local minimum value for the distance of
the orbits corresponding to $(\pi_{\phi}, f_{\pm}$ and
$(\pi_{\phi} + \delta \pi_{\phi}, f_{\pm} + \delta f_{\pm})$
equals
\begin{eqnarray}
\frac{1}{2}[ \delta f_{\pm}^2 + (1 + \frac{4
\pi_{\phi}^2}{|U'(\l_{1,2})|^2}) \delta \pi_{\phi}^2].
\end{eqnarray}
In the $U_<$ branch this  value is a global minimum, because at
$\l \rightarrow \infty$ the orbits diverge. In the $U_>$ branch,
however, the global minimum is achieved at $\l \rightarrow
\infty$, and corresponds to the metric of equation Eq.\
(\ref{redmetric}). Hence we conclude that the  metric in the
reduced state space has the value
\begin{eqnarray}
d\tilde{s}^2 = \frac{1}{2}[ d \tilde{f}_{\pm}^2 + (1 + \frac{4
\tilde{\pi}_{\phi}^2}{|U'(\l_{1}(\tilde{\pi}_{\phi}))|^2}) d
\tilde{\pi}_{\phi}^2], \label{redmetric2}
\end{eqnarray}
in the region $U_<$ (for re-collapsing universes), and the value
Eq.\ (\ref{redmetric}) in the region $U_>$.

There is clearly a discontinuity at the boundary region
$\pi_{\phi} ^2 = |U_{min}|$, where the metric Eq.\
(\ref{redmetric2}) diverges. This divergence is, however, an
artifact of the coordinates employed. The study of geodesics near
the boundary demonstrates that the distance function remains
finite.

Recall that at the boundary $\pi_{\phi} ^2 = |U_{min}|$, there
exist five different orbits for each value of $f-{\pm}$ and sign
of $\pi_{\phi}$. It is easy to verify that in the quantum theory
the distance function between those five orbits vanishes.
 Hence, all five orbits are described by one single coherent state in the quantum
 theory. It follows that the divergent points of the classical reduced
state space disappear in the quantum theory.  This is an
interesting result, because it suggests that the orbifold-like
structure of the reduced state space may be generically 'smeared',
when one passes to quantum theory.

Having identified the metric on the reduced state space, we employ
the path integral Eq.\ (\ref{pathintegral}) to construct the
overlap kernel on the reduced state space. For the cases $\kappa =
0, -1$ and for the regions $U_+, U_-, U_>$ of case $\kappa = 1$,
this yields the familiar Gaussian coherent states
\begin{eqnarray}
\hspace*{-0.2cm}\langle \tilde{\pi}_{\phi}, \tilde{f}_{\pm}|
\tilde{\pi}_{\phi}, \tilde{f}_{\pm} \rangle = \exp \!\left( i
\tilde{\pi}_{\phi} \tilde{f}_{\pm}' - i \tilde{\pi}_{\phi}'
\tilde{f}_{\pm} - \frac{1}{2} [ (\tilde{\pi}_{\phi} -
\tilde{\pi}_{\phi}')^2 + (\tilde{f}_{\pm} - \tilde{f}_{\pm}')^2]
\right).
\end{eqnarray}
For $\kappa = 1$ and an overlap kernel in which one of the entries
lies in $U_<$, we have to employ the metric Eq.\
(\ref{redmetric2}), from which it is very difficult to obtain an
analytic solution.

We should make a last comment, here, on the relation of this
procedure to the Dirac quantisation. The crucial difference in the
present method is that it does not quantise the constraints, but
it implements them at the classical level. Both algorithms are
expected to provide the same results in the semiclassical level.
However the reduced state space quantisation is not a
semiclassical approximation to the Dirac quantisation. It is a
quantisation method in its own right.

Whether the constraints are to be implemented at the quantum or
the classical level remains an open issue, that cannot be settled
{\em a priori}: both approaches yield a quantisation method that
provide the same classical limit. The choice between these
attitudes to canonical quantisation can only be determined by
their eventual success.

\section{Histories quantum R-W model}

\subsection{Representations of the Hilbert space}
The prescription for the implementation of the constraints
described in the previous section translates immediately to the
history context. The first step is to construct the Hilbert space
${\cal V}$, which carries a representation of the history group
with Lie algebra
\begin{eqnarray}
\[ \l_t, \pi_{\l t'} \] &=& i \delta (t,t') \\ \label{lpi}
\[ \phi_t, \pi_{\phi t'} \] &=& i \delta (t,t')\\ \label{phipi}
\[ N_t, \pi_{N t} \] &=& i \delta (t,t') \label{Npi}
\end{eqnarray}

The representation of the algebra Eqs.\ (\ref{lpi}-\ref{Npi}) is
selected by the choice of a canonical coherent states family, by
means of the inner product Eq.\ (\ref{innerprod}). From the
Gaussian coherent states Eq.\ (\ref{coherent}) we obtain
\begin{eqnarray}
 && \langle \l\!(\cdot),\pi_{\l}(\cdot),
\phi\!(\cdot), \pi_{\phi}\!(\cdot), N\!(\cdot),
\pi_N\!(\cdot)|\l'\!(\cdot),\pi_{\l}'\!(\cdot), \phi'\!(\cdot),
\pi_{\phi}'\!(\cdot), N'\!(\cdot), \pi_N'\!(\cdot) \rangle =
\nonumber
\\
 && \exp \left(\int\!\! dt[ i \pi_{\phi t} \phi'_t - i
\pi_{\phi t}' \phi_t + i \pi_{\l t} \l_t' - i \pi_{\l t}' \l_t  +i
\pi_{N t} N'_t - i \pi_{N t}' N_t \right.\\ \nonumber
 && \left. -\frac{1}{2}[ (\l_t - \l_t')^2 +
\frac{1}{w(t)}(\pi_{\l
t}- \pi_{\l t }')^2 + (\phi_t - \phi_t')^2 \right. \nonumber \\
&& \left. - \frac{1}{w(t)}(\pi_{\phi t} - \pi_{\phi t}')^2] - (N_t
- N_t')^2 - \frac{1}{w(t)}(\pi_{N t} -\pi_{N t}')^2] \right).
\label{coherenthis}
\end{eqnarray}

The function $w(t)$ is a density of weight $1$ and it is
introduced so that the definition of the integral is properly
invariant with respect to diffeomorphisms---see the discussion in
section 3.2. Different choices of $w$ lead to different
representations of the history group. The choice of $w$, however,
does not affect the probability assignment through the decoherence
functional and we shall, henceforward, set it to be equal to one.

Equation Eq.\ (\ref{coherenthis})  allows one to identify ${\cal
V}$ with the Fock space $e^{{\cal N}}$, where ${\cal N}$ is the
Hilbert space $L^2 (\R, C^3, dt)$. The history operators are
defined by means of the one parameter group of unitary
transformations they generate. For example, writing the smeared
operator $\hat{\l}(f) = \int dt \hat{\l}_t f(t)$, we define
\begin{eqnarray}
\hspace*{-0.2cm} e^{is \hat{\l}(f)}
|\l(\!\cdot\!),\pi_{\l}\!(\!\cdot\!), \phi(\!\cdot\!),
\pi_{\phi}\!(\!\cdot\!), N\!(\!\cdot\!), \pi_N\!(\!\cdot\!)
\rangle := |\l(\!\cdot\!),\pi_{\l}\!(\!\cdot\!) + s
f\!(\!\cdot\!), \phi(\!\cdot\!), \pi_{\phi}\!(\!\cdot\!),
N\!(\!\cdot\!), \pi_N\!(\!\cdot\!) \rangle
\end{eqnarray}
Next we construct the Hilbert space of the true degrees of freedom
${\cal V}_{phys}$. It is spanned by coherent state paths
$|\tilde{\pi}_{\phi} (\cdot), \tilde{f}_{\pm}(\cdot) \rangle$,
that are constructed from the  reduced coherent states
$|\tilde{\pi}_{\phi}, \tilde{f}_{\pm} \rangle$ of the standard
canonical theory.

Note that in the case $\kappa = 0, -1$ the reduced overlap kernel
 is Gaussian
\begin{eqnarray}
&& \langle \tilde{\pi}_{\phi}(\cdot), \tilde{f}_{\pm} (\cdot)|
\tilde{\pi}_{\phi}(\cdot), \tilde{f}_{\pm}(\cdot) \rangle = \\
\nonumber &&\exp \!\left(\int\!\! dt\;[ i \tilde{\pi}_{\phi t}
\tilde{f}_{\pm t}' - i \tilde{\pi}_{\phi t}' \tilde{f}_{\pm t}\! -
\!\frac{1}{2} [ (\tilde{\pi}_{\phi t}\! - \!\tilde{\pi}_{\phi
t}')^2 + (\tilde{f}_{\pm t }\! - \!\tilde{f}_{\pm t}')^2]]
\right).
\end{eqnarray}
The corresponding Hilbert space ${\cal V}_{phys}$ is a direct sum
of two copies of the Fock space $ e^{{\cal N}_{phys}}$, where
${\cal N}_{phys}$ is the Hilbert space $L^2(\R, C, dt)$. The one
copy corresponds to the expanding and the other to the collapsing
universe solutions.

\subsection{The decoherence functional}

We may also construct the decoherence functional on ${\cal
V}_{phys}$, which reads at the discrete level
\begin{eqnarray}
&& \hspace*{-1.3cm}d(\zeta_1,t_1; \zeta_2,t_2; \ldots ; \zeta_n,
t_n|\zeta'_1,t'_1;
\zeta'_2, t'_2; \ldots ; \zeta'_m,t'_m) =  \nonumber \\
&& \hspace*{-1.3cm}\langle \zeta'_m| \zeta_n \rangle \langle
\zeta_n| \zeta_{n-1} \rangle \ldots \langle \zeta_2| \zeta_1
\rangle \langle \zeta_1| \zeta_0 \rangle \langle
\zeta_0|\hat{\rho}_0|\zeta'_0 \rangle \times \langle \zeta'_0|
\zeta'_1 \rangle \langle \zeta'_1 | \zeta'_2 \rangle \ldots
\langle \zeta'_{n-1}| \zeta'_n \rangle.
\end{eqnarray}
At the continuous-time limit
\begin{eqnarray}
d[\zeta(\cdot), \zeta'(\cdot)] \sim e^{i\tilde{V}[\zeta(\cdot)] -
i\tilde{V}[\zeta'(\cdot)]},
\end{eqnarray}
where $\tilde{V}$ is the classical Liouville function  on the
reduced state space.

It is important to remark that the reduced state space
quantisation allows one to construct the decoherence functional
already at the level of the Hilbert space ${\cal V}$. The coherent
states construction induces a map may(?) employ the projection map
$\pi$ from the history constraint surface to the reduced state
space, in order to construct a map between the associated coherent
states
\begin{eqnarray}
i: |z(\cdot) \rangle \in {\cal V} \rightarrow |\zeta(\cdot)
\rangle \in {\cal V}_{his},
\end{eqnarray}
provided that $z$ is a path on the constraint surface. One may
then employ this map to {\em pull-back} the decoherence functional
on ${\cal V}$. Clearly the pull-backed decoherence functional has
support only on coherent state paths on the constraint surface.

We compare the above result for the decoherence functional, with
the one that results when applying the Dirac quantisation scheme.
To this end, we write a decoherence functional on ${\cal V}$ with
a delta function imposing the restriction of coherent state paths
on the constraint surface
\begin{eqnarray}
d(z_1,t_1; z_2,t_2; \ldots ; z_n, t_n|z'_1,t'_1;
z'_2, t'_2; \ldots ; z'_m,t'_m) =  \nonumber \\
  \langle z'_m| z_n \rangle \langle z_n|  z_{n-1} \rangle
\ldots \langle z_2| z_1 \rangle \langle z_1| z_0 \rangle \langle
z_0|\hat{\rho}_0 |z'_0 \rangle \nonumber \\
\times \langle z'_0| z'_1 \rangle \langle z'_1| z'_2 \rangle
\ldots \langle z'_{n-1}| z'_n \rangle \prod_{ij } \delta(C_{t_i})
\delta(C_{t'_j}),
\end{eqnarray}
where  $\delta(C_t)$ is the delta-function restricting paths on
the constraint surface.

Since the delta-function may be written as
\begin{eqnarray}
\delta(C) = \frac{1}{2 \pi} \int d \xi e^{i C \xi},
\end{eqnarray}
 then, at the continuous limit we obtain the following expression,
for the decoherence functional
\begin{eqnarray}
d[z(\cdot), z'(\cdot)] = \langle z_0|\hat{\rho}_0| z_0 \rangle
\int D \xi(\cdot) D \xi'(\cdot) e^{i (V - H_{\xi})[z(\cdot)]  - i
(V - H_{\xi'})[z'(\cdot)]}, \label{decfun3}
\end{eqnarray}
where $D \xi(\cdot) $ is the continuous limit of $\frac{d
\xi_{t_1} d \xi_{t_2} \ldots d \xi_{t_n}}{(2 \pi)^n}$, $H_{\xi} =
\int dt \xi(t) C_t$. Equation (\ref{decfun3})  which is a formal
expression analogous to equation Eq.\ (\ref{decfundirac}) obtained
from Dirac quantisation.

\subsection{Initial conditions and the classical limit}

If  the space of temporal supports $\cal{T}$ is real line, then we
need to restrict our considerations to coherent state paths
$\zeta(\cdot)$, that converge asymptotically to a fixed value
$\bar{\zeta}$, otherwise the inner product will be bounded.

For each value of $\bar{\zeta}$ we construct a different Hilbert
space and a different histories theory. We usually choose
$\bar{\z}$ by the requirement that $|\bar{\zeta} \rangle$ is the
minimum energy state. However, in the physical Hilbert space of
parameterised systems the Hamiltonian operator vanishes, therefore
we have no criterion for the selection of the vacuum state. A
different choice of the initial condition will yield a different
representation for the histories theory. Therefore, different
initial conditions define different theories, a fact that seems
singularly attractive at the cosmological context, which is the
only domain of physics in which the limit $t\rightarrow -\infty$
may be taken literally\footnote{We remind the reader that the
cosmological singularity lies outside the spacetime, so the
topology of the space of temporal supports in cosmology is
$(0,\infty)$ which is homeomorphic to the real line $\R$}.

It is easy to study the classical limit of the model, since the
Hamiltonian vanishes on the reduced state space. Omn\'es has
showed in\cite{Omn94, Omn89} that, for coarse-grained histories
that correspond to smearing within a phase space region of typical
size $L$, much larger than one ($\hbar = 1$), the off-diagonal
elements of the decoherence functional fall rapidly to zero. The
evolution peaks around the classical path with probability very
close to one\footnote{This is valid for a large class of
Hamiltonians including the trivial case of zero Hamiltonian.}.

When we apply the above result in the context of the
minisuperspace model problem, we notice that histories peaked
around the initial condition $|\bar{\zeta} \rangle$ have
probabilities very close to zero, hence, the system will be
adequately described by the corresponding classical solution.

It is rather interesting to provide an order of magnitude
estimation for processes that do not contribute to a classical
path, for example, processes that correspond to the various
proposals about the initial condition of the universe.

In the case of the minisuperspace model with $\kappa = +1$ we
consider, for example, an initial state within the region $U_<$
(corresponding to a universe with a bounded radius), and we
examine the probability of the realised history that corresponds
to an expanding universe.

We consider an initial state within $U_<$ with $\pi_{\phi}^2 =
\epsilon << 1$. This corresponds to a universe of radius $e^{\l}$,
close to zero. Since the point $e^{\l} = 0 $ is excluded from the
gravitational phase space (the metric is degenerate there) this is
the closer we can get to an initial condition analogous to that of
the tunnelling proposal for the wave function of the universe. For
simplicity, we shall assume that the initial state is
characterised by the value of $f_- = 0$---within the present model
this assumption does not affect the final results.

We may then ask about the probability that the universe will be
found sometime in the future within the region $U_<$, with
$\pi_{\phi} \simeq 0$, a state that corresponds to the onset of
inflation and evolve from there according to the classical
equations of motion.

We construct coarse-grained  histories, in which the reduced state
space is partitioned in cells, which are centered around specific
values for $\pi_{\phi}, f_{\pm}$, with typical length $L
>> 1$. The off-diagonal elements of the decoherence functional
between such histories will be very small (according to Omn\'es
theorem) and the probability that the universe will be found in a
cell $C$ centered around a specific value of $\pi_{\phi}, f_{\pm}$
will be approximately
\begin{eqnarray}
\int _C d \pi_{\phi} d f_{\pm} |\langle \epsilon, 0| \pi_{\phi},
f_{\pm} \rangle|^2 ( 1 + O(L^{-1})),
\end{eqnarray}
provided that the distance between the initial point $(\epsilon, 0
)$ and $\pi_{\phi}, f_{\pm}$ is much larger than $L$.

Next we need to calculate the quantity $|\langle \epsilon, 0|
\pi_{\phi}, f_{\pm} \rangle|^2$. We may estimate it taking into
account the following considerations. For a general family of
coherent states we write the norm of the overlap kernel $|\langle
z|z' \rangle|^2$
\begin{eqnarray}
|\langle z|z' \rangle|^2 = |\int   dz_1 dz_2 \ldots dz_n \langle
z|z_1 \rangle \langle z_2|z_3 \rangle \ldots \langle z_n|z'
\rangle|^2 \nonumber \\
< \int   dz_1 dz_2 \ldots dz_n |\langle z|z_1 \rangle \langle
z_2|z_3 \rangle \ldots \langle z_n|z' \rangle|^2.
\end{eqnarray}
For a large number of time-steps the above expression may be
expressed  as a path integral. At the continuous `time' limit we
may consider that $z_i' = z_i + \delta z_i$, in which case
$|\langle z_i|z_i + \delta z_i \rangle|^2 = e^{-2 \delta s_i^2}$,
in terms of the infinitesimal distance determined by the phase
space metric. Hence,
\begin{eqnarray}
|\langle z|z' \rangle|^2 < \int \prod dz_i e^{-2 \sum_i \delta
s_i^2} : = \int Dz(\cdot) e^{-2 L^2(z(\cdot))},
\end{eqnarray}
where the summation is over all paths $z(\cdot)$ joining $z$ and
$z'$, and $L(\cdot)$ is the length of the path $z(\cdot)$. Within
the saddle point approximation we may write then
\begin{eqnarray}
|\langle z|z' \rangle|^2 < c e^{- 2 D^2(z,z')}, \label{length}
\end{eqnarray}
where $c$ is a constant of order unity, and $D^2$ is the distance
function on the manifold, corresponding to the length of the
geodesic joining $z$ and $z'$.

For the study of the R-W minisuperspace model, we are especially
interested in the curve joining the point $(\epsilon, 0) \in U_<$
with the point $(0, f_-) \in U_>$. Since the regions $U_<$ and
$U_>$ are only connected at the surface $\pi_{\phi} = \pm
\sqrt{|U_{min}|}$, then,
\begin{eqnarray}
D^2[(\epsilon, 0), (0, f_-] = D^2[(\epsilon, 0),
(\sqrt{|U_{min}|}, v)] + D^2[(\sqrt{|U_{min}|}, v), (0, f_-)],
\end{eqnarray}
for some point on the surface $\pi_{\phi} = \pm \sqrt{|U_{min}|}$
characterised by the value $v$ of the coordinate  $f_{\pm}$.

The second distance can be calculated using the metric Eq.\
(\ref{redmetric}) and reads
\begin{eqnarray}
D^2[(\sqrt{|U_{min}|}, f), (0, f_-)] = \frac{1}{2}[|U_{min}| +
(f_- - v)^2].
\end{eqnarray}
The first distance concerns a path in $U_<$ and should be
calculated by using the metric Eq.\ (\ref{redmetric2}). However,
\begin{eqnarray}
 (1 + \frac{4 \pi_{\phi}^2}{|U'(\l_{1,2})|^2}) \delta \pi_{\phi}^2
\geq \delta \pi_{\phi}^2,
\end{eqnarray}
so that
\begin{eqnarray}
D^2[(\epsilon, 0), (\sqrt{|U_{min}|}, v)] > \frac{1}{2}[|U_{min}|
+ v^2],
\end{eqnarray}
whence
\begin{eqnarray}
D^2[(\epsilon, 0), (0, f_-]) > |U_{min}| + \frac{1}{2} [v^2 +
(v-f_-)^2] \geq |U_{min}| + f_-^2.
\end{eqnarray}
It follows that
\begin{eqnarray}
\langle \epsilon, 0| \pi_{\phi}, f_{\pm} \rangle|^2 < c
e^{-(2/3)^3/\Lambda^2 - f_-^2} < c e^{-(2/3)^3/\Lambda^2}.
\end{eqnarray}
Hence we have estimated an upper bound to the probability of
'tunnelling' from a universe of radius very close to one, to a
universe at the onset of inflation. Note that such a formula makes
sense only for coarse-grained histories at a phase space scale $L
>> 1$ and a distance $D^2$ on phase space much larger than $L$. It
follows that $D^2[(\epsilon, 0), (0, f_-])$ must be much larger
than one, hence $\Lambda^2 << 1$. Clearly, for a cosmological
constant of the order of the Planck scale ($\Lambda \sim 1$), it
would make no sense to discuss about tunnelling or a
 quasi-classical domain.

Our previous analysis has showed that the continuous-time
decoherence functional may be constructed with specific initial
states, in order to describe the scenario corresponding to the
tunnelling initial proposal of the universe.

An interesting issue arises, whether it would be possible to
extend the formalism to other initial conditions, most notably the
Hartle-Hawking proposal of no-boundary. The formalism presented
here cannot be immediately applied in this case. The reason is
that the Hartle-Hawking initial condition forces the consideration
of histories that do not correspond  to a Lorentzian
globally-hyperbolic metric, hence we do not have a straightforward
Hamiltonian analysis of constraints and time evolution.

However, the history formalism is sufficiently versatile to allow
for this scenario. The definition of histories only involves the
postulate of a partial ordering that implements the causality. It
is therefore applicable in principle to systems that have no
Hamiltonian description, either classically or quantum
mechanically . Such a generalisation should involve an enlarged
description of classical histories for general relativity, in
order to take into account the non-boundary four-metrics. More
importantly it requires the identification of a different
expression of the decoherence functional. Another, perhaps better
alternative would be a decoherence functional describing a quantum
growth process, similar to the one postulated in the causal set
scheme \cite{RiSo00}. This issue is at present under
investigation.

\section{Conclusions}

In this paper we studied the quantisation of minisuperspace models
within the framework of consistent histories. We showed that the
histories description preserves the notion of causal ordering at
the level of the true degrees of freedom, both classically and
quantum mechanically.

The key problem of minisuperspace models, compared to general
relativity, is that the imposition of the symmetry of homogeneity
and isotropy destroys the spacetime diffeomorphism symmetry of
general relativity. This leads to a number of problems at both the
classical and the quantum level.

Classically, one may retain some traces of the diffomorphism
symmetry at the level of the history reduced state space. However,
this involves the introduction of a metric-dependent foliation,
which is rather unnatural in the setting of the minisuperspace
model, unlike the case of general relativity.

At the quantum level, the loss of general covariance implies that
we have no reliable guide for the unique selection of a preferred
representation for the history Hilbert space.

We proposed a version of reduced state space quantisation for the
implementation of constraints, a procedure that was facilitated by
the key role played by coherent states in the construction of the
history Hilbert space. This procedure allows us to implement the
constraints at the quantum mechanical level, to construct the
decoherence functional, and thus to estimate probabilities for
interesting physical processes---such as the 'tunnelling' scenario
for the initial state of the universe.

An important result was that the quantisation procedure removes
from the physical Hilbert space all traces of the singular points
of the classical theory--the ones that render the reduced state
space into an orbifold.

\section*{Acknowledgements}
Research was supported by a Marie Curie Reintegration Grant of the
European Commission and a Pythagoras I grant from the Hellenic
Ministry of Education. Also, C. A. gratefully acknowledges
support from the Empirikion Foundation.

\end{document}